\documentclass[11pt, fullpage]{article}
\usepackage{fullpage}
\usepackage{setspace}
\usepackage{authblk}
\usepackage{cite}
\usepackage[colorlinks,citecolor=blue]{hyperref}
\usepackage{amsmath,amssymb,amsfonts}
\usepackage{algorithmic}
\usepackage{graphicx}
\usepackage{textcomp}

\usepackage[caption=false]{subfig}
\DeclareMathOperator*{\argmin}{arg\,min}  

\begin{document}

\title{An Accelerated Nonlinear Contrast Source Inversion Scheme for Sparse Electromagnetic Imaging}
\author[1]{Ali I. Sandhu}
\author[2]{Abdulla Desmal}
\author[1]{Hakan Bagci}
\affil[1]{Division of Computer, Electrical, and Mathematical Science and Engineering (CEMSE) \newline King Abdullah University of Science and Technology (KAUST), Thuwal, Saudi Arabia, 23955-6900 \newline \{aliimran.sandhu, hakan.bagci\}@kaust.edu.sa}
\affil[2]{Department of Electrical Engineering, Higher Colleges of Technology (HCT)\newline Ras Al-Khaimah, United Arab Emirates,  adesmal@hct.ac.ae}
\date{} 

\renewcommand\Affilfont{\itshape\small}

\maketitle
\onehalfspacing

\begin{abstract}
An efficient nonlinear contrast source inversion scheme for electromagnetic imaging of sparse two-dimensional investigation domains is proposed. To avoid generating a sequence of linear sparse optimization problems, the non-linearity is directly tackled using the nonlinear Landweber (NLW) iterations. A self-adaptive projected accelerated steepest descent (A-PASD) algorithm is incorporated to enhance the efficiency of the NLW iterations. The algorithm enforces the sparsity constraint by projecting the result of each steepest descent iteration into the $L_1$-norm ball and selects the largest-possible iteration step without sacrificing from convergence. Numerical results, which demonstrate the proposed scheme’s accuracy, efficiency, and applicability, are presented.
\vspace{0.5cm}


\end{abstract}
\newpage
\doublespacing

\section{Introduction}
\label{sec:introduction}
Electromagnetic (EM) imaging/inversion schemes are used to obtain (complex) permittivity profile in an investigation domain using scattered fields that are measured away from the investigation domain~\cite{pastorino2010microwave, pastorino2018microwave}. These schemes find applications in various fields including biomedical imaging~\cite{abubakar2002imaging, zhang2003microwave}, non-destructive testing~\cite{takagi1997electromagnetic}, crack detection~\cite{caorsi2001crack, pastorino2007detection}, subsurface imaging for hydrocarbon exploration~\cite{constable2007introduction}, and through-wall imaging~\cite{baranoski2008through}. Formulation and development of accurate and efficient EM inversion schemes face two fundamental challenges: (i) Scattered fields are a nonlinear function of the permittivity profile~\cite{pastorino2010microwave, pastorino2018microwave}. (ii) Scattered field measurements have to be carried out at a finite number of points and they are often contaminated with noise, rendering the EM inversion problem ill-posed~\cite{pastorino2010microwave,pastorino2018microwave}.

The strength of nonlinearity increases with the contrast level in the investigation domain. Here, the contrast level is defined as the relative difference between the permittivity in the investigation domain and permittivity of the background medium. For low contrast applications, first-order approximations to the nonlinear problem, such as the first-order Born and the Rytov approximations~\cite{rajan1989bornrytov}, provide accurate and efficient solutions. As the strength of the nonlinearity increases, higher-order approximations to the nonlinear scattering operator, such as the extended Born and the second-order Born schemes~\cite{estatico2005inexact} and the Born iterative method (BIM)~\cite{chew1989iterative}, have to be used to maintain the solution's accuracy and stability. For investigation domains with high contrast levels, fully nonlinear schemes, such as the distorted-Born method~\cite{chew1990reconstruction}, the inexact Newton method (INM)~\cite{bozza2009inexact}, and the nonlinear conjugate gradient scheme~\cite{abubakar2002imaging}, \cite{abubakar2005application}, have been developed for an accurate reconstruction of the investigation domain. Albeit effective, these schemes rely on local search strategies to find the solution of the nonlinear inverse problem. Therefore, the presence of local minima in the solution might negatively impact their convergence. On the other hand, stochastic optimization techniques~\cite{park1996microwave, price2013differential, robinson2004pso} can be used to obtain the global minima but this comes at the cost of significantly increased computational requirements.

To circumvent the ill-posedness of the EM inversion that is often cast in the form of a minimization problem, regularization schemes are used~\cite{pastorino2010microwave, pastorino2018microwave}. These schemes often use prior information about the unknown permittivity profile to constrain the minimization problem, which shrinks the solution space and yields a more stable and more accurate approximation to the solution. Until recently, $L_2$-norm of the solution has been the traditional choice as the constraint term in EM inversion algorithms~\cite{chew1989iterative, chew1990reconstruction, franchois1997microwave, bozza2009inexact}. These algorithms converge to a smooth solution rather efficiently, but their accuracy or convergence deteriorates when they are applied to the investigation domains with sharp variations/discontinuities and/or sparse content in their contrast profile. Note that the contrast profile may exhibit spatial sparseness (meaning that it is nonzero only in a small portion of the investigation domain), or it might have a sparse representation in another domain such as those obtained using wavelet transform~\cite{winters2010sparsity}, discrete cosine transform \cite{fornasier2010theoretical}, and spatial derivatives~\cite{sandhu2016bim}. Recently, it has been demonstrated that using $L_0 / L_1$ - norm of the solution as the constraint (instead or together with the $L_2$-norm) results in more accurate and efficient reconstructions for the investigation domains with sharp variations/discontinuities and/or sparse content~\cite{winters2010sparsity, bagci2008sparsity, desmal2014shrinkage, desmal2015preconditioned, sandhu2016bim}. It is well known that solving an $L_0$-constraint optimization problem is not feasible as it has NP-time complexity; however greedy algorithms developed under the compressed sensing paradigm have been shown to provide a good approximation to the solution of the $L_0$-constrained EM inversion problem~\cite{pan2012compressive, csenyuva2015electromagnetic}. Alternatively, convex relaxation-based regularization schemes, where the $L_1$-norm constrained problem is considered as the best approximation to the $L_0$-constrained problem, are widely used. In~\cite{poli2012microwave, anselmi2018iterative}, a Bayesian compressed sensing approach has been implemented to solve the EM inversion problem under the first-order Born approximation. In~\cite{oliveri2014compressive} a total variation based compressed sensing framework has been shown to promote sharp edges in piecewise continuous contrast profiles. These schemes are either computationally intensive, or their accuracy is limited since a first-order approximation to the nonlinear scattering operator is used. Note that this paper focuses on a deterministic EM inversion scheme, and therefore the readers are referred~\cite{ poli2012microwave, anselmi2018iterative} and the references therein for more details on the Bayesian frameworks used for EM inversion.

In~\cite{desmal2014shrinkage} and~\cite{desmal2015preconditioned}, the nonlinear scattering operator is linearized using the BIM~\cite{chew1989iterative} and the INM~\cite{bozza2009inexact} respectively. The linear system obtained at every iteration of these methods is then solved via the Landweber (LW) iterations. To enforce the sparsity constraint ($L_0 / L_1$ - norm of the solution),  the well-known shrinkage thresholding functions~\cite{daubechies2004iterative, blumensath2009iterative} are applied to the result of every LW iteration. In~\cite{desmal2015sparse}, a scheme that makes use of nonlinear Landweber (NLW) iterations has been developed to directly solve the $L_0$/$L_1$-norm constrained nonlinear minimization problem. The sparsity-constraint is again enforced using thresholding functions applied during the NLW iterations. Unlike the INM scheme in~\cite{desmal2015preconditioned}, the NLW iterations do not call for generation and solution of a sequence of linear sparse optimization problems (i.e., tackle the non-linearity directly) and require selection/tweaking of a smaller set of parameters~\cite{li2008image}. On the other hand, the NLW iterations, both the traditional version used for $L_2$ - norm-constrained regularization and the thresholded version, often suffer from slow convergence which prohibits their direct application to electrically large investigation domains.

To overcome this bottleneck of slow convergence, a projected accelerated steepest descent (PASD) algorithm has been used to solve the three-dimensional (3D) EM inverse scattering problem~\cite{desmal2017pasd, daubechies2008acc_proj_gradient, teschke2010accelerated}. To enforce the sparsity constraint, this algorithm projects the result of each steepest descent iteration into the $L_1$-norm ball. The steepest descent iterations are faster than NLW iterations since they allow for selection of a larger iteration step without sacrificing from accuracy and convergence. Consequently, the resulting PASD algorithm achieves significant speed up over the thresholded NLW iterations. However, even with the increased rate of convergence, the computational cost of every iteration is still a limiting factor in applying the PASD scheme to electrically large investigation domains. This high computational cost is a result of the matrix inversions required to compute the forward operator and its (first-order) Frechet derivative~\cite{desmal2017pasd}.

To this end, in this work, the overall computational cost of the nonlinear EM inversion is further reduced by incorporating a self-adaptive version of the PASD algorithm~\cite{han2016self_adap_projec} and the nonlinear contrast source (CS) formulation of EM scattering~\cite{van2001contrast, van1999extended_CSI}. The self-adaptive PASD scheme (A-PASD) chooses the largest possible iteration step in a recursive manner (with smaller number of operations) and increases the convergence of the original PASD algorithm. The CS formulation uses both the contrast and equivalent currents as the unknowns to be solved for in the nonlinear optimization. Even though the total number of unknowns is increased, the forward operators and its Frechet derivative do not call for matrix inversions. The resulting A-PASD-CS scheme benefits from the faster convergence rate of the A-PASD algorithm and the lower computational cost (per iterations step) that comes with the CS formulation.    
  
The proposed A-PASD-CS scheme is used for solving the nonlinear EM inverse scattering problem on 2D (spatially) sparse investigation domains. Numerical results demonstrate that the proposed scheme is more accurate and efficient than the existing CS-based EM inversion schemes. 

\section{Formulation}
\subsection{Contrast Source Formulation} \label{sec:cs_formulation}
\begin{figure}[!t]
\centering
	{\includegraphics[width=0.50\columnwidth]{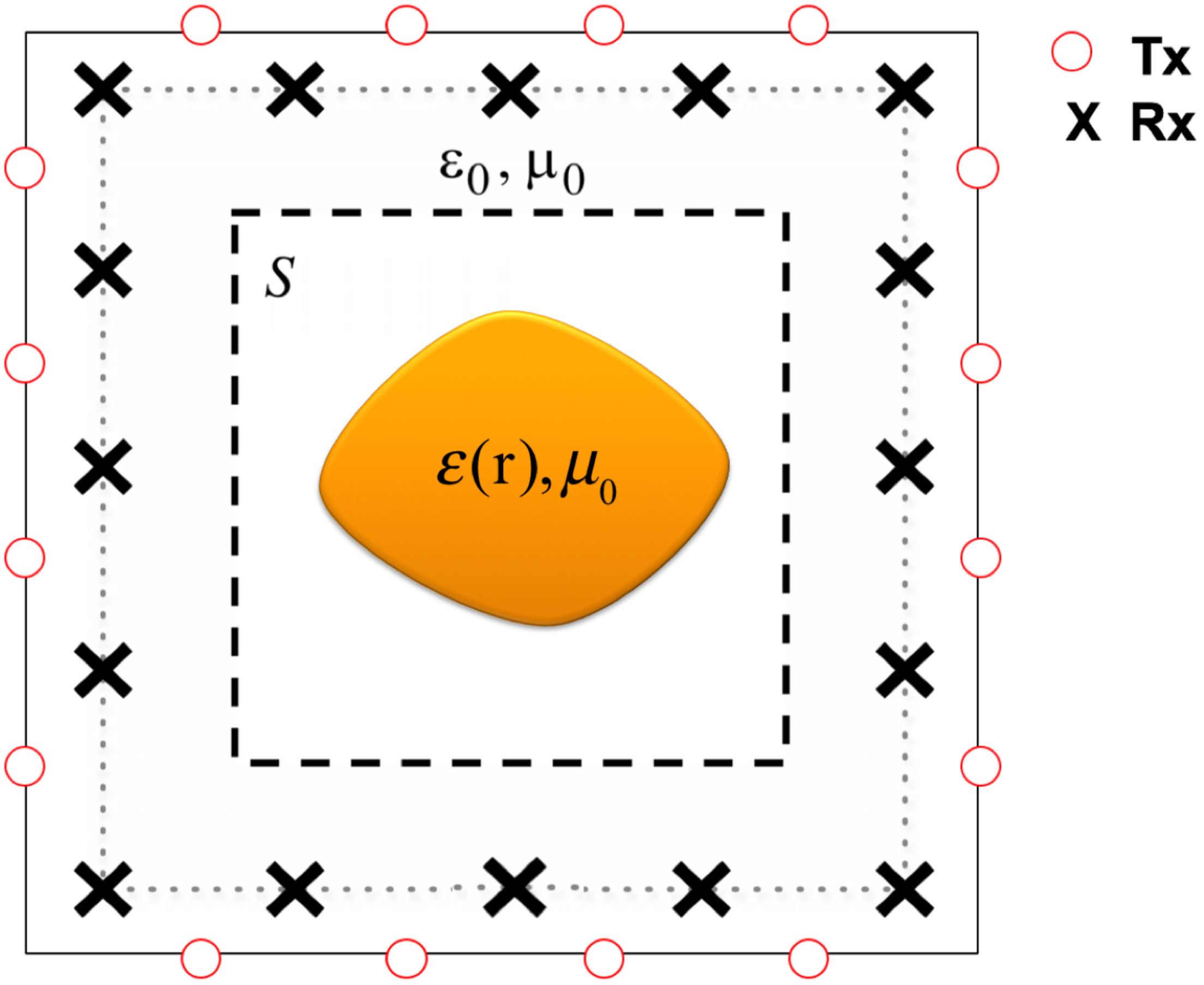}}
	\caption{Description of the 2D EM inversion problem.}
	\label{fig1}
\end{figure}

Let $S$ represent a 2D investigation domain embedded in an unbounded homogeneous medium. The permittivities and permeabilities in $S$ and the background medium are $\{ \varepsilon ({\mathbf{r}}),{\mu _0}\} $ and $\{ {\varepsilon _0},{\mu _0}\} $, respectively. $S$ is surrounded by ${N^{\mathrm{T}}}$ number of line sources (Fig. \ref{fig1}). It is assumed that $S$ is excited by these sources individually, i.e., by one of them at a time. Let $E_i^{{\mathrm{inc}}}({\mathbf{r}})$, $i = 1,..,{N^{\mathrm{T}}}$, represent the transverse-magnetic (to $z$), ${\rm TM}^{z}$,  polarized electric field, generated by the ${i^{{\mathrm{th}}}}$ source. Upon excitation, equivalent volume electric current ${J_i}({\mathbf{r}})$ is induced in $S$ and ${J_i}({\mathbf{r}})$ generates the scattered electric field $E_i^{{\mathrm{sca}}}({\mathbf{r}})$, which is expressed as
\begin{equation} \label{eq1}
T_i^{\mathrm{R}}({J_i}) = E_i^{{\mathrm{sca}}}({\mathbf{r}}) = {k_0^2}\int\limits_S {{J_i}({\mathbf{r'}})G({\mathbf{r}},{\mathbf{r'}})ds'}.
\end{equation} 
Here, $G({\mathbf{r}},{\mathbf{r'}}) = H_0^2({k_0}\left| {{\mathbf{r}} - {\mathbf{r'}}} \right|)/(4j)$ is the 2D scalar Green function, ${k_0} = \omega \sqrt {{\varepsilon _0}{\mu _0}} $ is the wavenumber in the background medium, and $\omega $ is the frequency. Additionally, ${J_i}({\mathbf{r}})$ is related to the total electric field ${E_i}({\mathbf{r}})$ through ${J_i}({\mathbf{r}}) = \tau ({\mathbf{r}}){E_i}({\mathbf{r}})$, where $\tau(\mathbf{r}) =\varepsilon(\mathbf{r})/\varepsilon_0-1$ is the contrast. Substituting \eqref{eq1} for $E_i^{{\mathrm{sca}}}({\mathbf{r}})$ into the fundamental field relation ${E_i}({\mathbf{r}}) = E_i^{{\mathrm{inc}}}({\mathbf{r}}) + E_i^{{\mathrm{sca}}}({\mathbf{r}})$, ${\mathbf{r}} \in S$, and multiplying the resulting equation with $\tau ({\mathbf{r}})$ yield:
\begin{align}
\label{eq2} T_i^{\mathrm{S}}(\tau ,{J_i}) & = {J_i}({\mathbf{r}}) - \tau ({\mathbf{r}})E_i^{{\mathrm{inc}}}({\mathbf{r}}) 
 - k_0^2\tau({\mathbf{r}})\int\limits_S {{J_i}({\mathbf{r'}})} G({\mathbf{r}},{\mathbf{r'}})ds' = 0, \, {\mathbf{r}} \in S.
\end{align}
The coupled set of equations \eqref{eq1} and \eqref{eq2} are known as the CS formulation~\cite{van2001contrast, van1999extended_CSI}. Let $E_i^{{\mathrm{mea}}}({\mathbf{r}}_m^{\mathrm{R}})$, $m = 1, \ldots ,{N^{\mathrm{R}}}$, represent the measured values of $E_i^{{\mathrm{sca}}}({\mathbf{r}})$, where ${\mathbf{r}}_m^{\mathrm{R}}$, $m = 1, \ldots ,{N^{\mathrm{R}}}$ are the measurement locations. Then, the CS equations  \eqref{eq1} and \eqref{eq2} can be numerically solved for unknowns ${J_i}({\mathbf{r}})$ and $\tau ({\mathbf{r}})$ given $E_i^{{\mathrm{inc}}}({\mathbf{r}})$ and $E_i^{{\mathrm{mea}}}({\mathbf{r}}_m^{\mathrm{R}})$, $m = 1, \ldots ,{N^{\mathrm{R}}}$ as described in the next two sections.
\subsection{Discretization} \label{sec:discretization}
To facilitate the numerical solution, first, $S$ is discretized into $N$ number of square elements. Let $\mathbf{r}_n$, $n = 1, \ldots ,N$ represent the centers of these elements. Then, the unknowns ${J_i}({\mathbf{r}})$ and $\tau ({\mathbf{r}})$ are approximated as 
\begin{align}\label{eq3}
\nonumber {J_i}({\mathbf{r}}) = \sum\limits_{n = 1}^N {{{\{ {{\bar J}_i}\} }_n}{p_n}({\mathbf{r}})}\\ 
\tau ({\mathbf{r}}) = \sum\limits_{n = 1}^N {{{\{ \bar \tau \} }_n}{p_n}({\mathbf{r}})} 	
\end{align}
where ${\{ {\bar J_i}\} _n} = {J_i}({{\mathbf{r}}_n})$, ${\{ \bar \tau \} _n} = \tau ({{\mathbf{r}}_n})$, and ${p_n}({\mathbf{r}})$ is the pulse basis function on element $n$ with support ${S_n}$. ${p_n}({\mathbf{r}})$ is non-zero only for ${\mathbf{r}} \in {S_n}$ with unit amplitude. Substituting the first expansion in \eqref{eq3} into \eqref{eq1} and evaluating the resulting equations at ${\mathbf{r}}_m^{\mathrm{R}}$, $m = 1, \ldots ,{N^{\mathrm{R}}}$ yield: 
\begin{equation}\label{eq4}        
\bar T_i^{\mathrm{R}}({\bar J_i}) = \bar G_i^{\mathrm{R}}{\bar J_i} = \bar E_i^{{\mathrm{mea}}}
\end{equation} 
where 
\begin{equation}{
\nonumber\{ \bar G_i^{\mathrm{R}}\} _{m,n}} = k_0^2\int_{{S_n}} {G({\mathbf{r}}_m^{\mathrm{R}},{\mathbf{r'}})ds'}
\end{equation}
and ${\{ \bar E_i^{{\mathrm{mea}}}\} _m} = E_i^{{\mathrm{mea}}}({\mathbf{r}}_m^{\mathrm{R}})$. Substituting \eqref{eq3} into \eqref{eq2} and evaluating the resulting equations at ${{\mathbf{r}}_m}$, $m = 1, \ldots ,N$ yield: 
\begin{equation}\label{eq5} 
\bar T_i^{\mathrm {S}}(\bar \tau ,{\bar J_i}) = {\bar J_i} - D[\bar \tau ]\bar E_i^{{\mathrm{inc}}} - D[\bar \tau ]\bar G_i^{\mathrm{S}}{\bar J_i} = {\bar 0}
\end{equation}  
where 
\begin{equation}
\nonumber {\{ \bar G_i^{\mathrm{S}}\} _{m,n}} = k_0^2\int_{{S_n}} {G({{\mathbf{r}}_m},{\mathbf{r'}})ds'}
\end{equation}
${\{ \bar E_i^{{\mathrm{inc}}}\} _m} = E_i^{{\mathrm{inc}}}({{\mathbf{r}}_m})$, and operator $D[.]$ generates a diagonal matrix with entries equal to the entries of the vector at its argument. Equations \eqref{eq4} and \eqref{eq5} are cascaded together for all excitations $i = 1,..,{N^{\mathrm{T}}}$ in a more compact form as: 
\begin{equation}\label{eq6} 
\bar T(\bar z) - \bar y = {\bar 0}
\end{equation}
where 
\begin{align}
\nonumber \bar T(\bar z)  = & [\bar T_1^{\mathrm{S}}(\bar \tau,{\bar J_1}),\ldots,\bar T_{N^{\mathrm{T}}}^{\mathrm{S}}(\bar \tau ,{\bar J_{N^{ \mathrm {T}}}}),\bar T_1^{\mathrm{R}}({\bar J_1}),\dots, \bar T_{N^{\mathrm{T}}}^{\mathrm{R}}({\bar J_{N^{\mathrm{T}}}})]^{t}\\
\nonumber \bar z = &{[{\bar \tau ^{t}},\bar J_1^{t},\bar J_2^{t}, \ldots ,\bar J_{{N^{\mathrm{T}}}}^{t}]^{t}}\\
\nonumber \bar y = & {[{\bar 0^{t}}, \ldots ,{\bar 0^{t}} ,\bar E_1^{{\mathrm{mea}},t},\bar E_2^{{\mathrm{mea}},t}, \ldots ,\bar E_{{N^{\mathrm{T}}}}^{{\mathrm{mea}},t}]^{t}}.
\end{align}
Here, superscript `$t$' represents the transpose operation. 
\subsection{Sparsity Regularized Nonlinear Optimization Problem} \label{sec:nonlinear optimization}
Equation \eqref{eq6} is nonlinear in $\bar z$ but also ill-posed and cannot be solved accurately/efficiently without using a regularization method~\cite{pastorino2010microwave,pastorino2018microwave}. In this work, it is assumed that the investigation domain is sparse, i.e., many entries of $\bar \tau $, and consequently the same entries of ${\bar J_i}$,  $i = 1,..,{N^{\mathrm{T}}}$,  are zero.  Therefore, to alleviate the ill-posedness, \eqref{eq6} is cast in the form of a sparsity-constrained nonlinear optimization problem: 
\begin{equation}\label{eq7}  
\bar z^{*} = \mathop {\argmin} \limits_{ {\bar z} } \frac{1}{2}\left\| {\bar y - \bar{T}( \bar z )} \right\|_2^2,\quad {\left\|  \bar z \right\|_0} \leq {l_0}.  
\end{equation}
In \eqref{eq7}, the nonlinear least squares minimization, i.e., the data misfit $\left\| {\bar y - \bar T(\bar z)} \right\|_2^2$ is constrained by the condition ${\left\| {\bar z} \right\|_0} \leq {l_0}$, which ensures that the total number of non-zero entries in $\bar z$ (given by ${\left\| {\bar z} \right\|_0}$), is less than the positive integer ${l_0}$. The solution of \eqref{eq7} provides the sparsest result possible, however, since it is a non-convex optimization problem, this solution is NP-hard and computationally not feasible~\cite{daubechies2004iterative, massa2015compressive}. The best convex approximation to \eqref{eq7} is obtained by constraining the data misfit with the ${L_1}$-norm of $\bar z$: 
\begin{equation}\label{eq8} 
\bar z^{*} = \mathop {\argmin} \limits_{ {\bar z} } \frac{1}{2}\left\| {\bar y - \bar T( {\bar z} )} \right\|_2^2,\quad {\left\| { {\bar z} } \right\|_1} \leq {l_1}. 
\end{equation}
In \eqref{eq8}, ${l_1}$ is a positive real number that should be estimated based on the prior knowledge of the sparsity level of $\bar z$ and the values of its entries. A plethora of algorithms has been developed to solve the ${L_1}$- norm constrained nonlinear optimization problem~\cite{daubechies2004iterative,blumensath2009iterative,bioucas2007new}. In this work, the scheme of nonlinear Landweber iterations (NLW) is selected from this group of algorithms to solve \eqref{eq8}. At every iteration of this scheme, a thresholding function is applied to enforce the sparsity constraint. The NLW scheme is preferred here because it requires a smaller number of parameters to be ``tuned'' heuristically in comparison to the inexact Newton and Born iterative methods that have been previously developed for solving the inverse scattering problem on sparse domains~\cite{desmal2015preconditioned,desmal2014shrinkage}. However, it has also been shown that the classical NLW scheme converges fast at the beginning of the iterations, but then it overshoots the ${L_1}$-norm penalty and takes a large number of iterations to correct back~\cite{daubechies2008acc_proj_gradient}. 

\subsection{Self-Adaptive Projected Accelerated Steepest Descent Scheme} \label{sec:PASD}
In this work, the slow convergence of the NLW scheme is alleviated using a self-adaptive version of the PASD algorithm~\cite{teschke2010accelerated, han2016self_adap_projec}. The original PASD scheme achieves acceleration by confining the solution search within a particular $L_{1}$-norm ball, while maintaining a large iteration step size without sacrificing from accuracy and convergence~\cite{teschke2010accelerated}. Its self-adaptive version, which is proposed in~\cite{han2016self_adap_projec} and abbreviated as A-PASD in this paper, further increases the convergence by controlling the step size in a recursive/adaptive manner. This approach starts with a larger step size and decreases it only when necessary. The application of the A-PASD to the solution of the nonlinear problem \eqref{eq8} yields in the following algorithm:
\begin{align*}
&{\mathrm{Step}}\;1:\; {\mathrm{Initialize}}\; l_{1}, \alpha, \; \psi, \; \mu \in (0,1), \; \delta \in (0,1)
\rho \in (0,1), \; \lambda_{(k\ge1)}>0, \; \gamma_{(0)}=\beta_{(0)} > 0, \; {\mathrm{ and}} \; \bar{z}_{(0)} \\ 
&{\mathrm{Step}}\;2:\; {\bold{for}}\; k=0,1,2,.... \\
&{\mathrm{Step}}\;2.1: \quad p_{k} = 0\\
&{\mathrm{Step}}\; 2.2: \quad {\bold{do}}\; {\mathrm{until}}\; \bar{z}_{(k+1)}\; {\mathrm{satisfies}}\; {\mathrm{condition}}\; (9)\\
&{\mathrm{Step}}\; 2.2.1:\quad\quad p_{k} = p_{k} + 1\; {\mathrm{ and}}\; \beta_{(k+1)} = \mu^{p_{(k)}}\gamma_{(k)} \\
&{\mathrm{Step}}\; 2.2.2:\quad\quad {\mathrm{find}}\; {\mathrm{ fixed}}\; {\mathrm{ point}}\; {\mathrm{ of}}\\
&\quad\quad \bar{z}_{(k+1)} = P \big(\bar{z}_{(k)} +\frac{\beta _{(k+1)} }{r} \partial_{\bar{z}_{(k+1)} } \bar{T}^\dag \left(\bar{y}-\bar{T}(\bar{z}_{(k)} )\right)\big)\\
&{\quad\qquad\quad\quad}\;{\bold{ end}}\; {\bold{ do}} \\
&{\mathrm{Step}}\; 2.3:\quad {\bold{ if}}\; \bar{z}_{(k+1)}\; {\mathrm{and}}\; \beta_{(k+1)}\; \mathrm{satisfy}\; {\mathrm{ condition}}\; (10) \\
&\quad\quad\quad\quad\quad\quad\quad \gamma_{(k+1)} = (1+\lambda_{(k+1)})\beta_{(k+1)}\\
&\quad\quad\quad\quad\quad\; {\bold{ else} }\\
&\quad\quad\quad\quad\quad\quad\quad \gamma_{(k+1)} = \beta_{(k+1)}\\
&\quad\quad\quad\quad\quad\; {\bold{ end}}\; {\bold{ if}}\\
&\quad\quad\quad\quad {\bold{ end}}\; {\bold{ for}}\\
\end{align*}
At Step 1, several parameters are initialized: $\alpha$ and $\psi$ should be selected to satisfy $\alpha$ $\geq {\sup _{\bar z \in B}}$ ${\left\| {{\partial _{\bar z}}\bar T(\bar z)} \right\|_2}$ and $\psi  \geq {\sup _{\{ \bar z,\bar u\}  \in B}}{{2{{\left\| {\partial _{\bar z}^2\bar T(\bar z,\bar u)} \right\|}_2}} \mathord{\left/{\vphantom {{2{{\left\| {\partial _{\bar z}^2\bar T(\bar z,\bar u)} \right\|}_2}} {\left\| {\bar u} \right\|_2^2}}} \right. \kern-\nulldelimiterspace} {\left\| {\bar u} \right\|_2^2}}$, where ${\partial _{\bar z}}\bar T(\bar z)$ and $\partial _{\bar z}^2\bar T(\bar z,\bar u)$ are the first- and second-order Frechet derivatives~\cite{desmal2015preconditioned} and $B = \{ {\left\| {\bar z} \right\|_1} \leq {l_1}\} $  is a ball of radius ${l_1}$~\cite{teschke2010accelerated}. 

Parameters $\mu$, $\gamma_{(k)}$, and $\lambda_{(k)}$ help to reduce the number of ``trials'' in finding the iteration step size $\beta_{(k)}/r$ (see Steps 2.2.1, 2.2.2, and 2.3)~\cite{han2016self_adap_projec}. The parameter $r=\max \left\{2\alpha ,2\psi \sqrt{\Gamma ( \bar{z}_{(0)} ) } \right\}$, where $\Gamma (\bar z) = 0.5\left\| {\bar y - \bar T(\bar z)} \right\|_2^2$ is the data misfit, and helps in adjusting the iteration step size [see Step 2.2.2, \eqref{eq9}, and \eqref{eq10}]. Parameters  $\delta$ and $\rho$, which are on the right hand sides of \eqref{eq9} and \eqref{eq10}, respectively, help in ensuring the convergence of the algorithm. 

Step 2 and its sub-steps describe the iterations. ``do loop'' that starts at Step 2.2 adjusts the iteration step size in an adaptive/recursive manner until the condition~\cite{han2016self_adap_projec, ge2011global_stepsize, he2002modified_glp_projection} 
\begin{equation}\label{eq9}
\left\| \bar{T}( \bar{z}_{(k + 1)} )- \bar{T}( \bar{z}_{(k)} ) \right\|^{2} \leq \frac{(1 - \delta)r}{\beta_{(k+1)}} \left\| \bar{z}_{(k + 1)} - \bar{z}_{(k)} \right\|^2 \!
\end{equation}
is satisfied. 

Step 2.2.2 is the fixed point iteration of the solution, where $P( \cdot )$ is the projection operator and ${\partial _{\bar z}}{\bar T^\dag }( \cdot )$ is the Hermitian transpose of ${\partial _{\bar z}}\bar T( \cdot )$. It is important to note here that $\bar z_{(k)}$ contain samples of the contrast $\tau({\mathbf{r}})$ and equivalent currents ${J_i}({\mathbf{r}})$, which have values that are orders of magnitude different from each other. The effect of this scaling mismatch is observed in the singular values of ${\partial _{\bar z}}{\bar T^\dag }( \cdot )$, and consequently, the iterations converge very slowly. In this work, a left-right diagonal preconditioning scheme (that is similar to the one used in \cite{desmal2015preconditioned}) is applied at Step 2.2.2 to alleviate the effect of this scaling mismatch and increase the convergence rate.  

The purpose of Step 2.3 is to increase the iteration step size if it gets too small. This is checked by the condition  
\begin{equation}\label{eq10} 
\left\| \bar{T}( \bar{z}_{(k + 1)} ) - \bar{T}( \bar{z}_{(k)} ) \right\|^{2} \leq \frac{\rho r}{\beta_{(k+1)}} \left\| \bar{z}_{(k + 1)}-\bar{z}_{(k)}  \right\|^2. \!\!
\end{equation}
Note that there could be several positive sequences satisfying $\lambda_{(k\ge1)}>0$. In this work, $\lambda_{(k)} = \lambda_0$, where $\lambda_0$ is a positive constant.

In the above algorithm, the projection operator $P( \cdot )$ is computed using~\cite{desmal2017pasd} 
\begin{align*}
&{\mathrm{Step}}\,1:\bar x = {\mathrm{sort}}\left( {\left| {\bar z} \right|} \right) \\
&{\mathrm{Step}}\,2:{\mathrm{ Find}}\, m\, {\mathrm{  s.t.}}\, {\left\| {\mathrm{Thr}}^{{\{\bar x\}}_m}(\bar x) \right\|_1} \!\! \leq \! l_1\! \leq \!\! {\left\| {\mathrm{  Thr}}^{{\{\bar x\}}_{m+1}}(\bar x) \right\|_1} \\
&{\mathrm{Step}}\,3:\chi = \{ \bar x\}_m-\left(l_1-\left\| {\mathrm{  Thr}}^{\{ \bar x\}_m}(\bar x)\right\|_1\right)/m \\
&{\mathrm{Step}}\,4:P(\bar z) = {\mathrm{  Thr}}^\chi(\bar z) 
\end{align*}

At Step 1, vector $\bar x$ stores the absolute of value of the entries of the input vector $\bar z$ sorted in descending order. Steps 2 and 3 are needed to compute the threshold level used at Step 4. Here, ${\mathrm {Thr}}^\chi( \cdot )$ is the complex soft-thresholding function defined as~\cite{desmal2017pasd} 
\[{\{ {\mathrm{Th}}{{\mathrm{r}}^\chi }(\bar z)\} _m} = {\{ \bar z\} _m}\frac{{\max \left\{ {\left| {{{\{ \bar z\} }_m} - \chi } \right|,0} \right\}}}{{\max \left\{ {\left| {{{\{ \bar z\} }_m} - \chi } \right|,0} \right\} + \chi }}\] 
where $\chi$ is the thresholding level.

Note that the mathematical derivation and justification of the conditions, steps, and projection operation used by the A-PASD algorithm are not provided in here but can be found in~\cite{teschke2010accelerated, han2016self_adap_projec, daubechies2008acc_proj_gradient}. In this work, this algorithm is applied to accelerate the solution of the EM inverse problem constructed using the CS formulation. The numerical results presented in the next section demonstrate that the resulting inversion scheme is indeed faster and more accurate than existing CS-based EM inversion schemes.

\section {Numerical results}  \label{sec:numerical_results}

This section demonstrates the accuracy and efficiency of the proposed scheme via numerical experiments. In all examples considered here, $E_{i}^{\mathrm{mea}}(\mathbf{r}_{m}^{\mathrm{R}})$, $m=1,\ldots ,{{N}^{\mathrm{R}}}$, $i=1,..,{{N}^{\mathrm{T}}}$, are generated synthetically: First, \eqref{eq5} with ${{\bar{\tau }}^{\mathrm{ref}}}$ is solved for ${{\bar{J}}_{i}}$, then ${{\bar{J}}_{i}}$ is inserted into \eqref{eq4}, and finally $25\mathrm{dB}$ Gaussian noise is added to the result to yield $E_{i}^{\mathrm{mea}}(\mathbf{r}_{m}^{\mathrm{R}})$.  Here, ${{\{{{\bar{\tau }}^{\mathrm{ref}}}\}}_{n}}={{\tau }^{\mathrm{ref}}}({{\mathbf{r}}_{n}})$, $n=1,\ldots ,N^{\mathrm{ref}}$, are the samples of the actual contrast ${{\tau }^{\mathrm{ref}}}(\mathbf{r})$.

Unless otherwise stated, in all examples, $f=125$ MHz (corresponding to a wavelength of $\lambda = 2.4\,\mathrm{m}$), ${N}^{\mathrm{T}} = 8$  and  ${N}^{\mathrm{R}} = 16$, and the investigation domain is of dimensions $7.5\,\mathrm{m} \times 7.5\,\mathrm{m}$ ($3.125\lambda \times 3.125\lambda$). To discretize ${{\tau }^{\mathrm{ref}}}(\mathbf{r})$, $N^{\mathrm{ref}}=3600$, which corresponds to an element length of $0.125\,\mathrm{m}=\lambda/19.2$. To discretize $\tau(\mathbf{r})$ (contrast being reconstructed), $N^{\mathrm{ref}}=2500$, which corresponds to an element length of $\Delta d = 0.15\, \mathrm{m}=\lambda/16$. Note that the mesh used for discretizing ${{\tau }^{\mathrm{ref}}}(\mathbf{r})$ is denser than the one used for discretizing $\tau(\mathbf{r})$ to avoid the issue of "inverse crime''  \cite{colton1998inverse}.

Three different EM inversion schemes are compared: (i) A preconditioned inexact Newton scheme that incorporates the CS formulation and thresholded (linearized) LW iterations~\cite{desmal2015preconditioned}. This scheme is referred to as SP-IN-CS. (ii) A multi-resolution scheme with the CS formulation. This scheme uses a coarse discretization at the beginning of the simulation to recover the general structure of the solution and as the simulation evolves, finer details are reconstructed using finer discretizations. This approach is referred to as MR-CS~\cite{barth2001multiscale}. (iii) A-PASD-CS algorithm proposed in this work. 

The parameters of the A-PASD-CS algorithm are provided in Table~\ref{tab1}. As explained in Section~\ref{sec:cs_formulation}, parameters $\alpha$ and $\psi$ have to satisfy conditions that depend on the suprema of the Frechet derivatives in ball $B$. These Frechet derivatives are functions of the problem setup (transmitter-receiver configuration, discretization, frequency, scatterer and domain size, etc.) and their suprema (and therefore $\alpha$ and $\psi$ are determined heuristically by running some numerical tests on the given problem before the A-PASD-CS scheme is executed~\cite{teschke2010accelerated, han2016self_adap_projec}. Parameters $\rho$, $\delta$, and $\lambda_0$ control step size of the A-PASD algorithm. Their combination provided in Table~\ref{tab1}, which ensures a good convergence rate for the problems considered in this section, is also determined heuristically by running several numerical tests~\cite{teschke2010accelerated, han2016self_adap_projec}.

Moreover, as shown in Table~\ref{tab1}, that for the three problems considered in this section, two but all the parameters remained similar. This demonstrates that, once these parameters are optimized, only a small tuning is needed to solve inversion problems with similar setups. This reduces the overall dimension of the configuration parameters and consequently keep the calibration complexity similar to the existing-state-of-the-art sparsity regularized inversion schemes.

For all simulations, the quality of reconstruction is measured using
\begin{equation} 
err_t=\frac{{{\left\| {{{\bar{\tau }}}_{t}}-{{{\bar{\tau }}}^{\mathrm{ref}}} \right\|}_{2}}}{{{\left\| {{{\bar{\tau }}}^{\mathrm{ref}}} \right\|}_{2}}}
\end{equation}  
where ${{\bar{\tau }}_{t}}$ stores the samples of the contrast reconstructed at the iteration corresponding to the execution time $t$. All simulations are carried on a $3.5$ GHz $64$-Core Intel Xeon E$5$ processor with $64\, \mathrm{GB}$ RAM.

\subsection{Coaxial}
\begin{figure*}[!h]
	\centering
	\subfloat[]{\includegraphics[width=0.25\columnwidth]{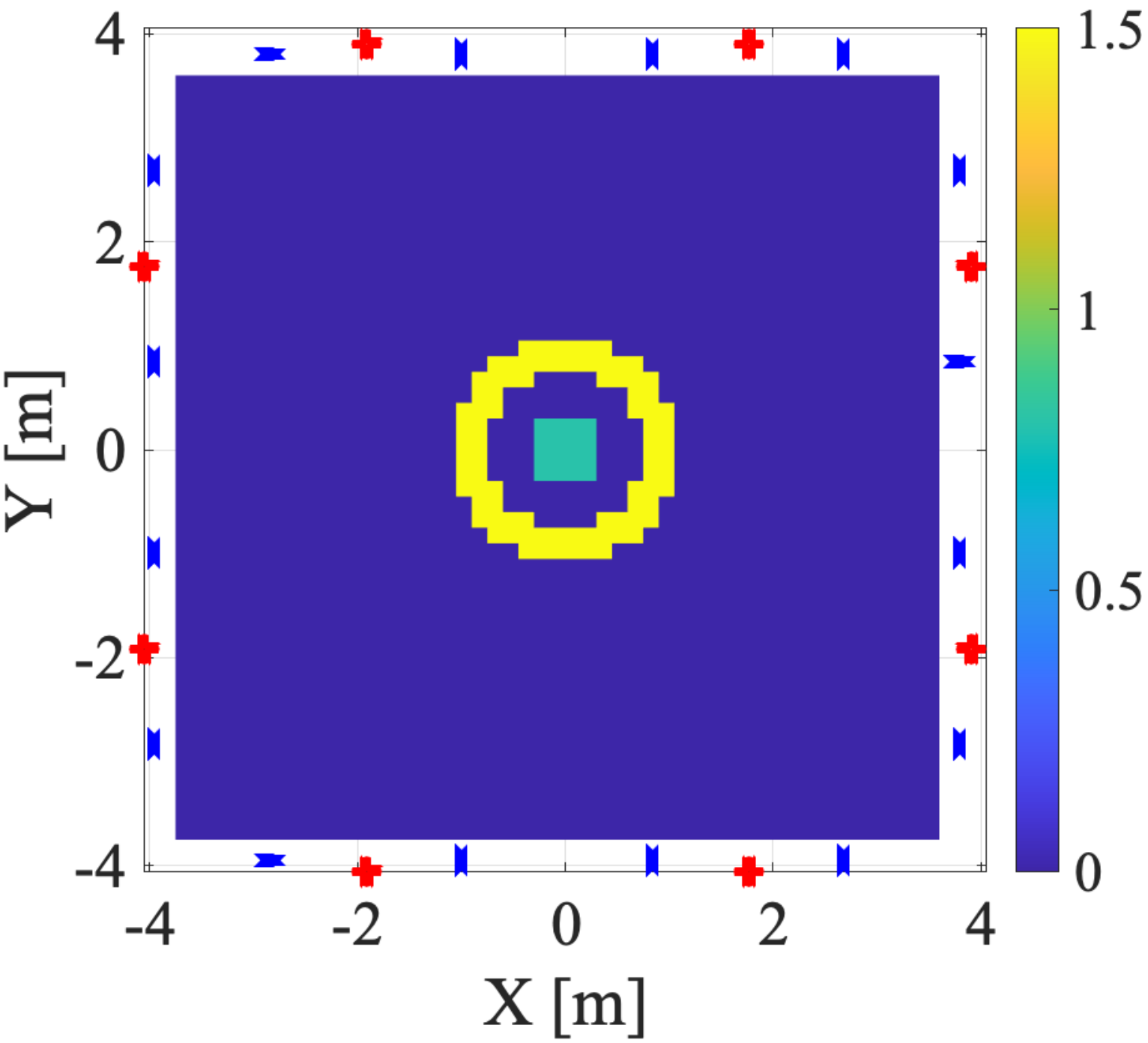}}
	\subfloat[]{\includegraphics[width=0.24\columnwidth]{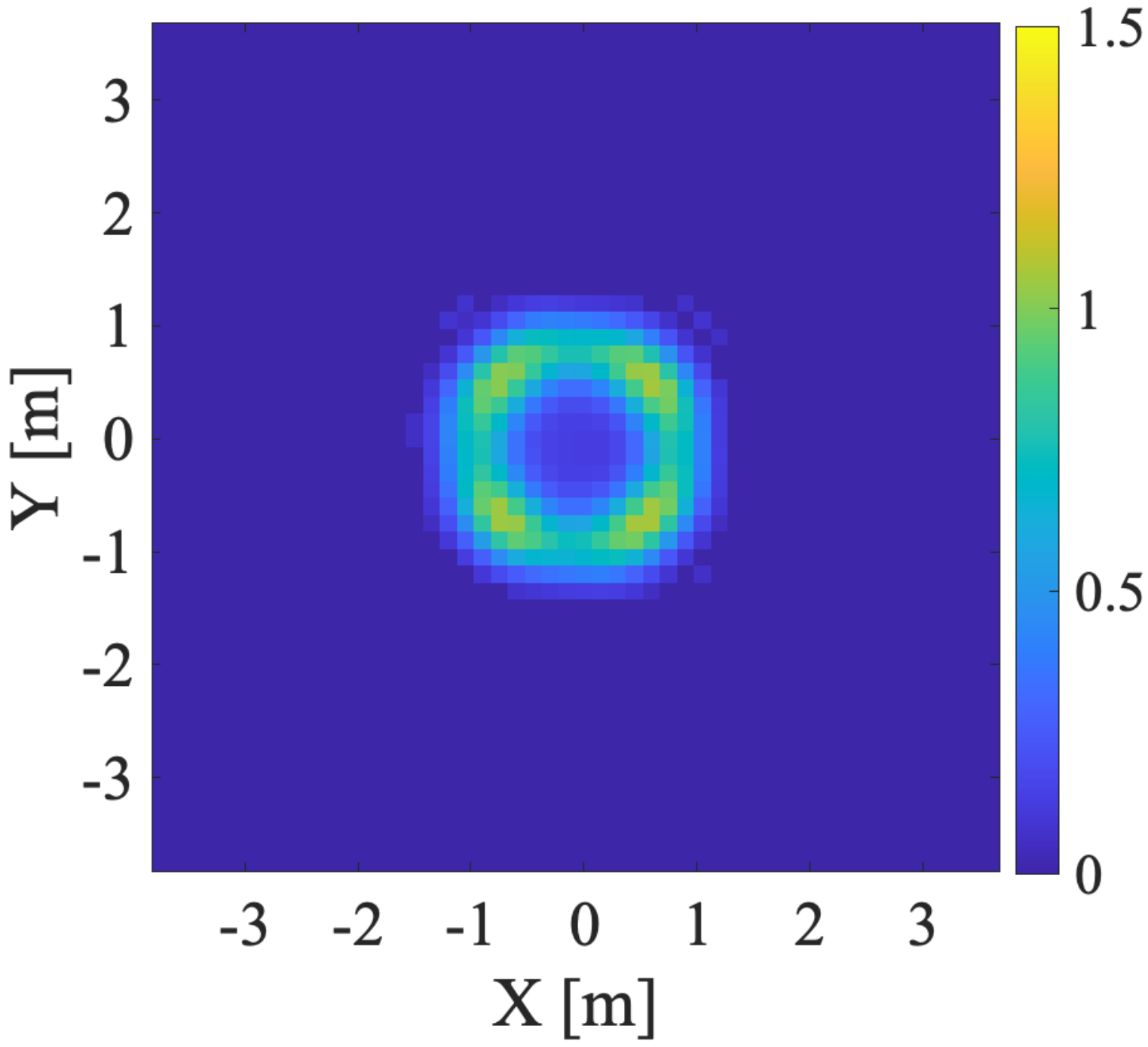}}\\
	\subfloat[]{\includegraphics[width=0.24\columnwidth]{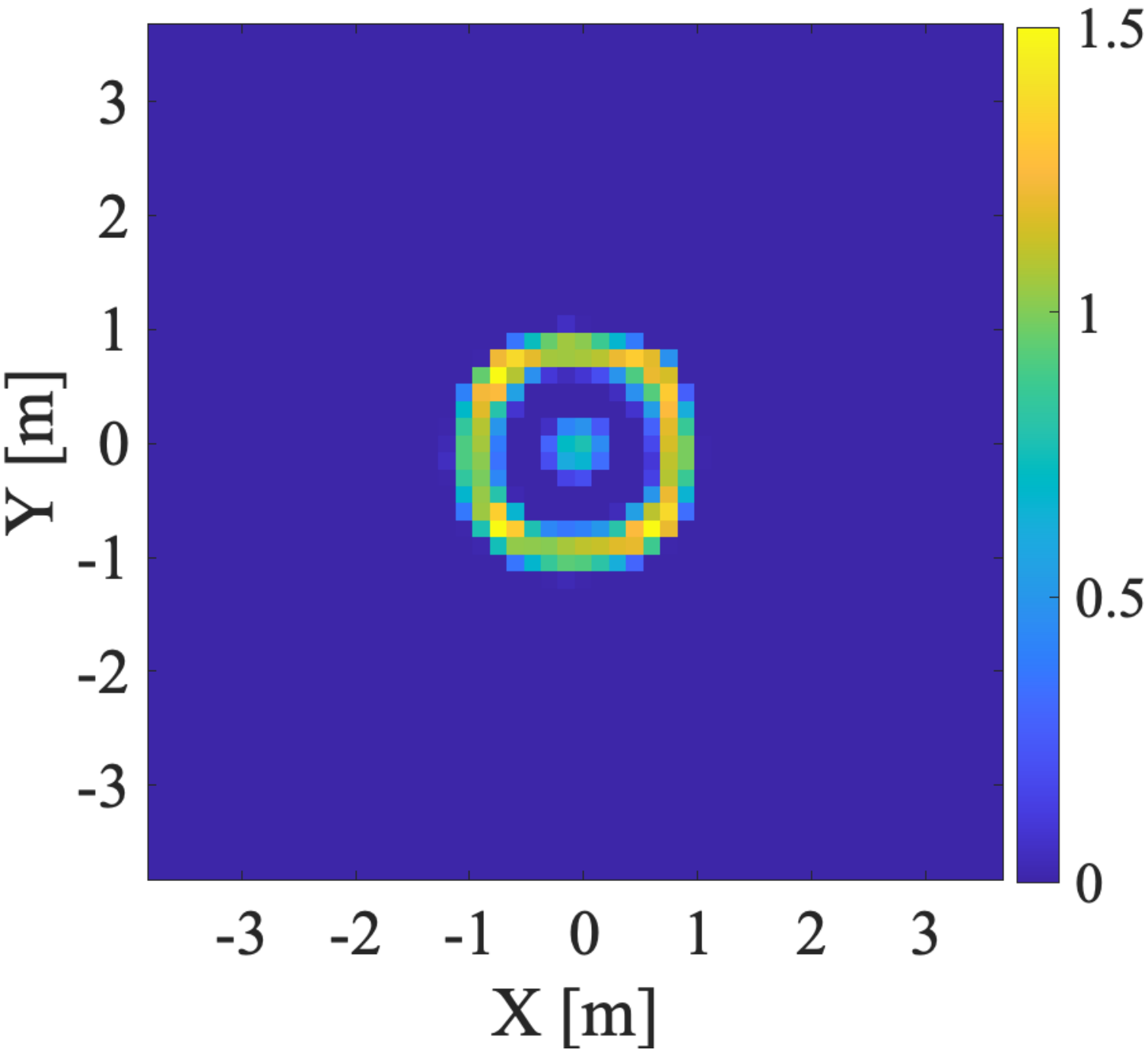}}
	\subfloat[]{\includegraphics[width=0.24\columnwidth]{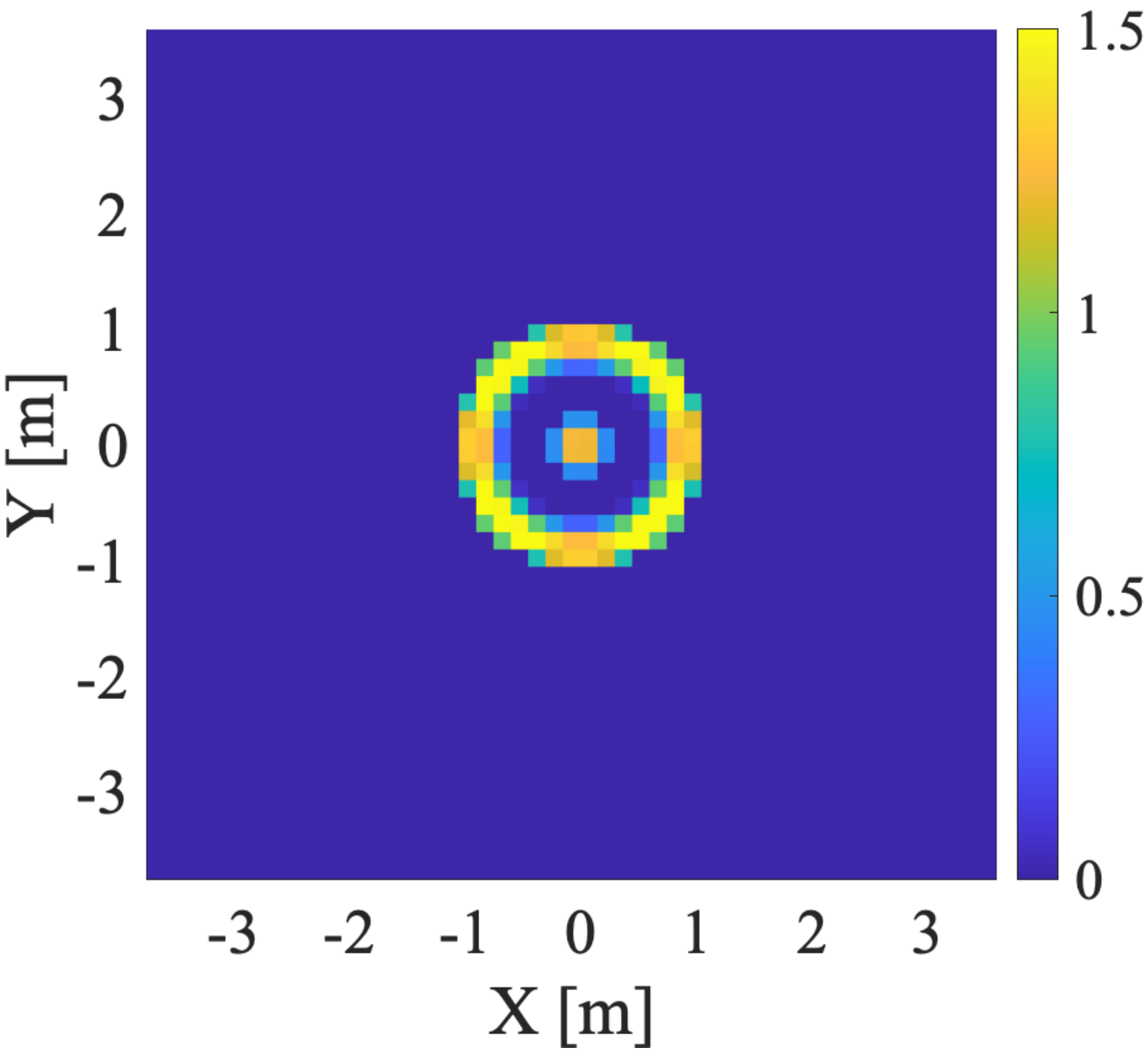}} \\
	\subfloat[]{\includegraphics[width=0.50\columnwidth]{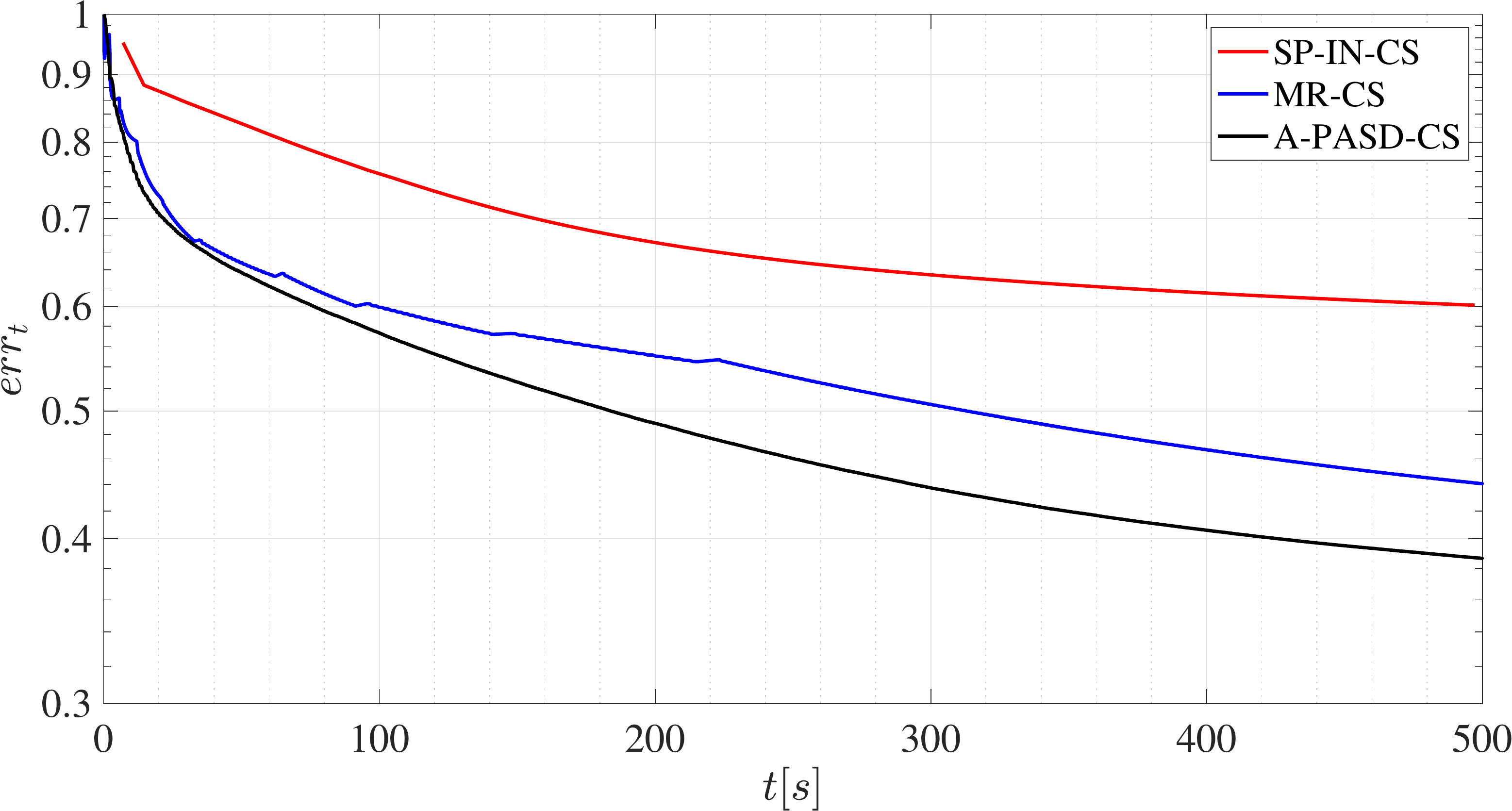}} \\
	\subfloat[]{\includegraphics[width=0.50\columnwidth]{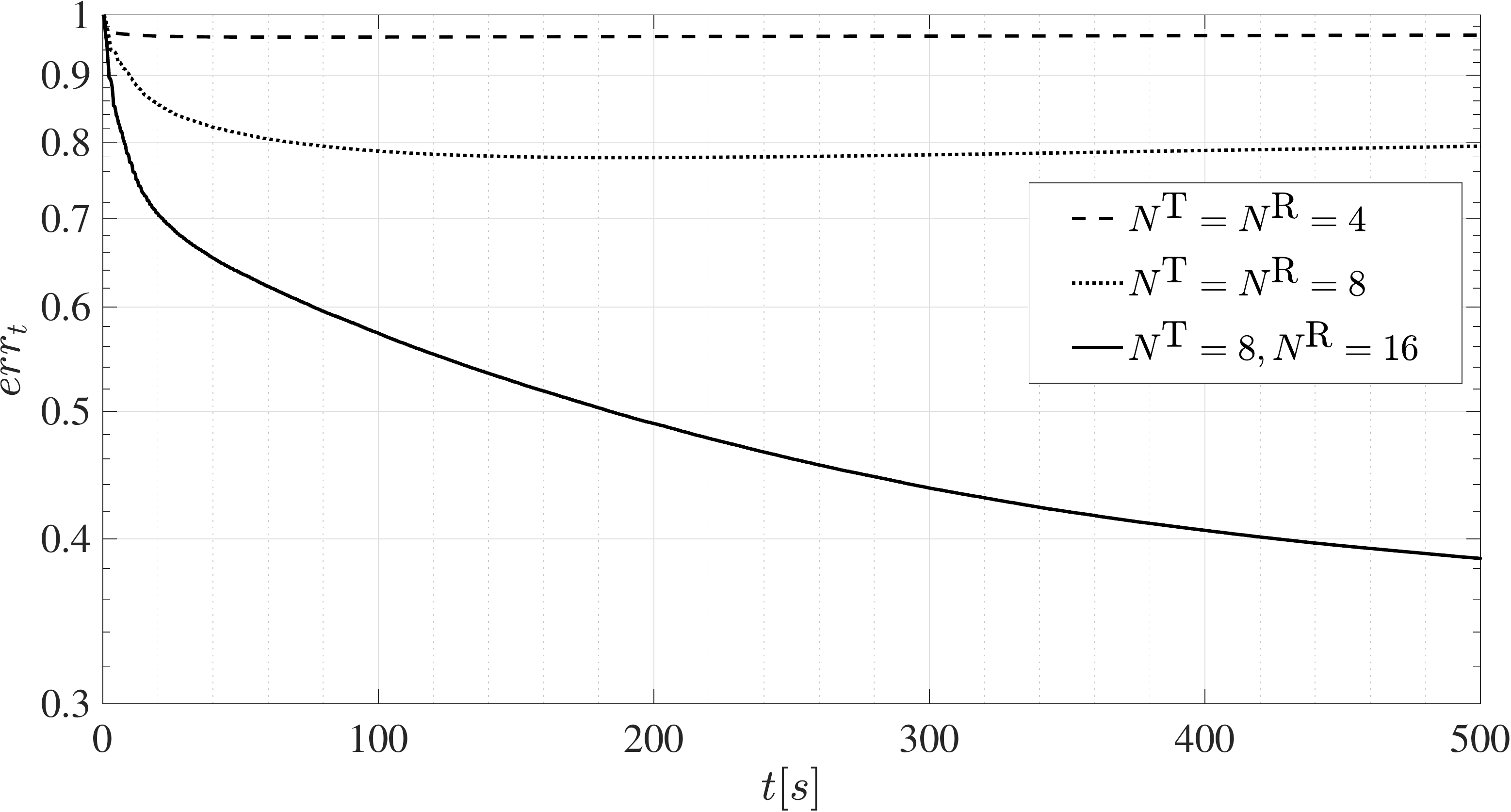}}
	\caption{(a) Investigation domain with the coaxial-shaped scatterer (as represented by $\bar{\tau}^{\mathrm{ref}}$) and the transmitter and receiver locations. Solutions obtained by (b) SP-IN-CS, (c) MR-CS, and (d) A-PASD-CS. (e) Reconstruction error $err_t$ versus execution time $t$ for all three schemes. (f) Reconstruction error $err_t$ versus execution time $t$ for A-PASD-CS with different ${N}^{\mathrm{T}}$ and ${N}^{\mathrm{R}}$.}
\label{fig2}
\end{figure*}

The first example is of a ``coaxial''-shaped scatter. The outer ring and the inner cylinder are centered at the origin. Outer and inner radii of the ring are $1.2\,\mathrm{m}$ and $0.9\,\mathrm{m}$, respectively. The radius of the inner cylinder is $0.45\,\mathrm{m}$. The relative permittivities of the outer ring and the inner cylinder are $1.8$ and $2.5$, respectively. The investigation domain (as represented by ${{\bar{\tau }}^{\mathrm{ref}}}$) and transmitter-receiver configuration are shown in Fig.~\ref{fig2}(a). The sparseness level (the ratio of number of non-zero entries in ${{\bar{\tau }}^{\mathrm{ref}}}$ to $N$) is $4\%$. 

Fig.~\ref{fig2} (b)-(d) show ${\bar \tau}_{t=504 \, {\mathrm{s}}}$, ${\bar \tau}_{t=501 \, {\mathrm{s}}}$, and ${\bar \tau}_{t=501 \, {\mathrm{s}}}$ obtained using SP-IN-CS, MR-CS, and A-PASD-CS, respectively. The corresponding error levels are  $err_{t=504 \, {\mathrm{s}}}=60\%$, $err_{t=501 \, {\mathrm{s}}}=44\%$ and $err_{t=501 \, {\mathrm{s}}}=38\%$. Note that the number of iterations required by SP-IN-CS, MR-CS, and A-PASD-CS to produce these images is $72$, $7905$, and $7830$, respectively. Fig. \ref{fig2}(e) plots $err_t$ (in log scale) versus execution time $t$ for all three schemes. These results show that A-PASD-CS converges faster than the other two schemes and produces a more accurate reconstruction at the point of convergence. 

Next, the effect of ${N}^{\mathrm{T}}$ and ${N}^{\mathrm{R}}$ on the accuracy of the A-PASD-CS scheme is demonstrated. Different sets of transmitter-receiver configurations, where ${N}^{\mathrm{T}} \in \{4,8\}$ and ${N}^{\mathrm{R}} \in \{4,8,16\}$, are considered. In all sets, transmitters and receivers are uniformly located around the investigation domain. Fig.\ref{fig2}(f) plots $err_t$ versus execution time $t$ for various combinations of $\{{N}^{\mathrm{T}}, {N}^{\mathrm{R}}\}$. As expected, the reconstruction accuracy and the convergence rate increase with the increasing number of measurements ${N}^{\mathrm{T}}\times{N}^{\mathrm{R}}$.

\subsection{Austria}
\begin{figure}[!t]
	\centering
	\subfloat[\label{}]{\includegraphics[width=0.25\columnwidth]{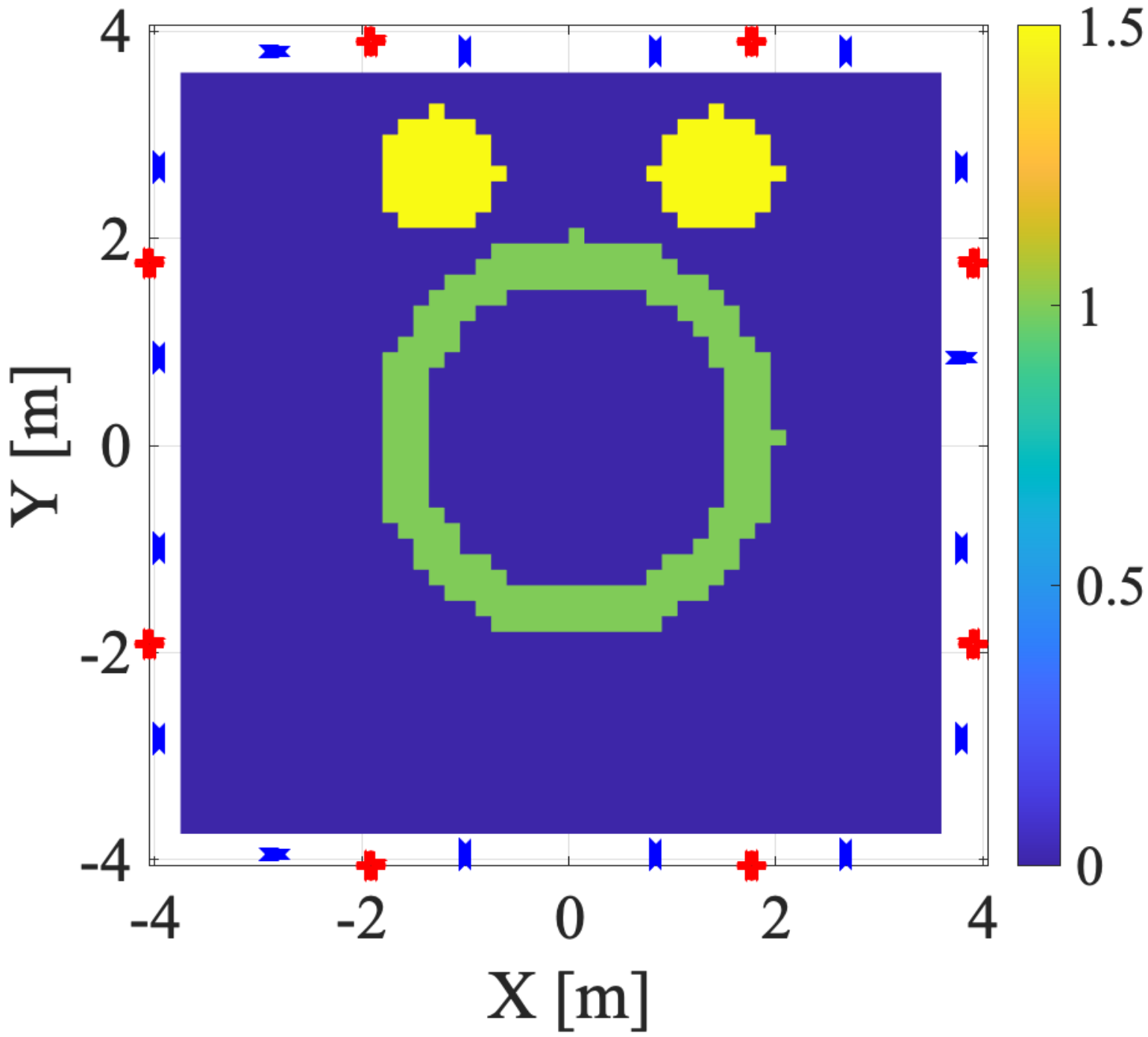}}
	\subfloat[\label{}]{\includegraphics[width=0.24\columnwidth]{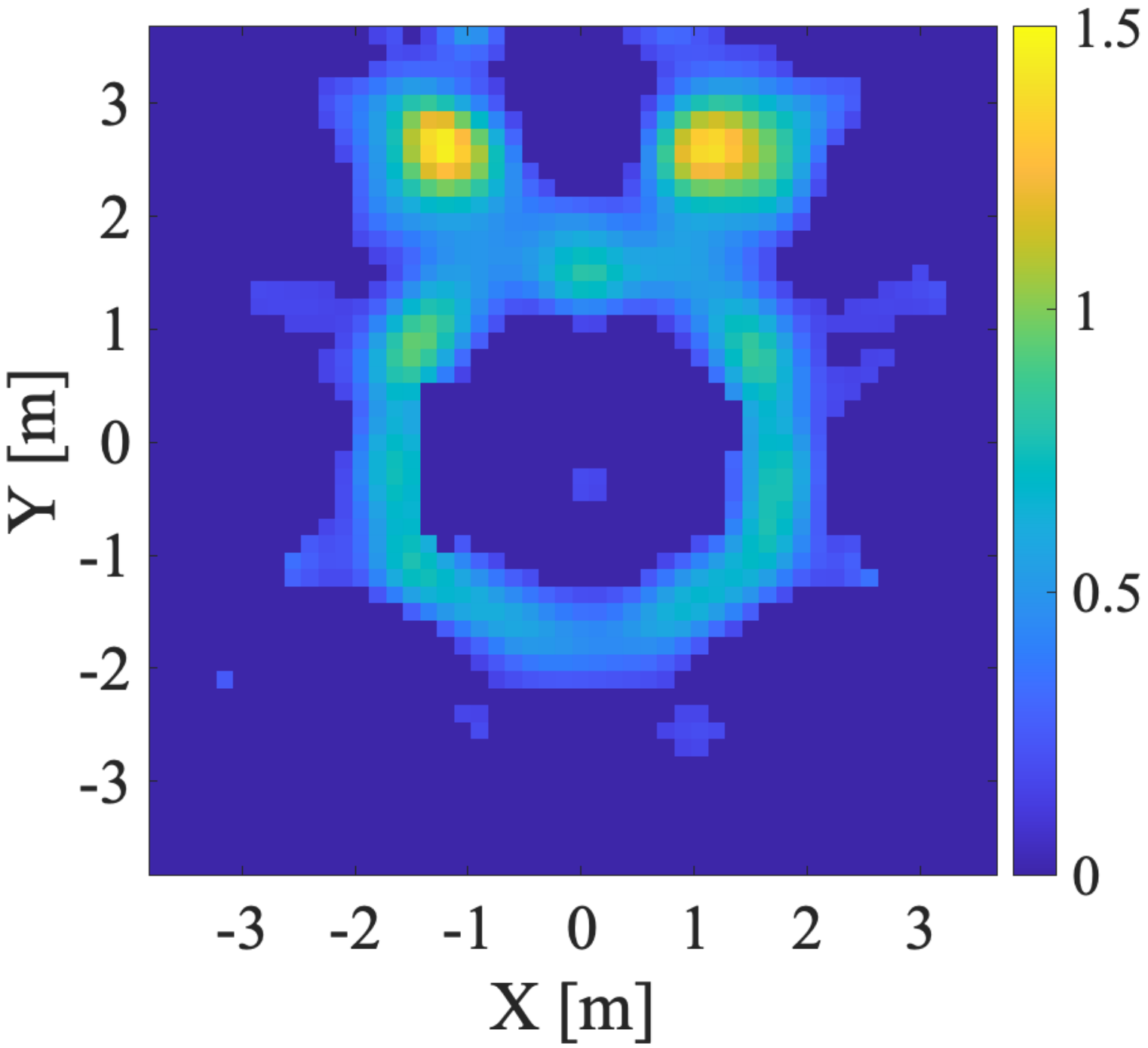}} \\
	\subfloat[\label{}]{\includegraphics[width=0.24\columnwidth]{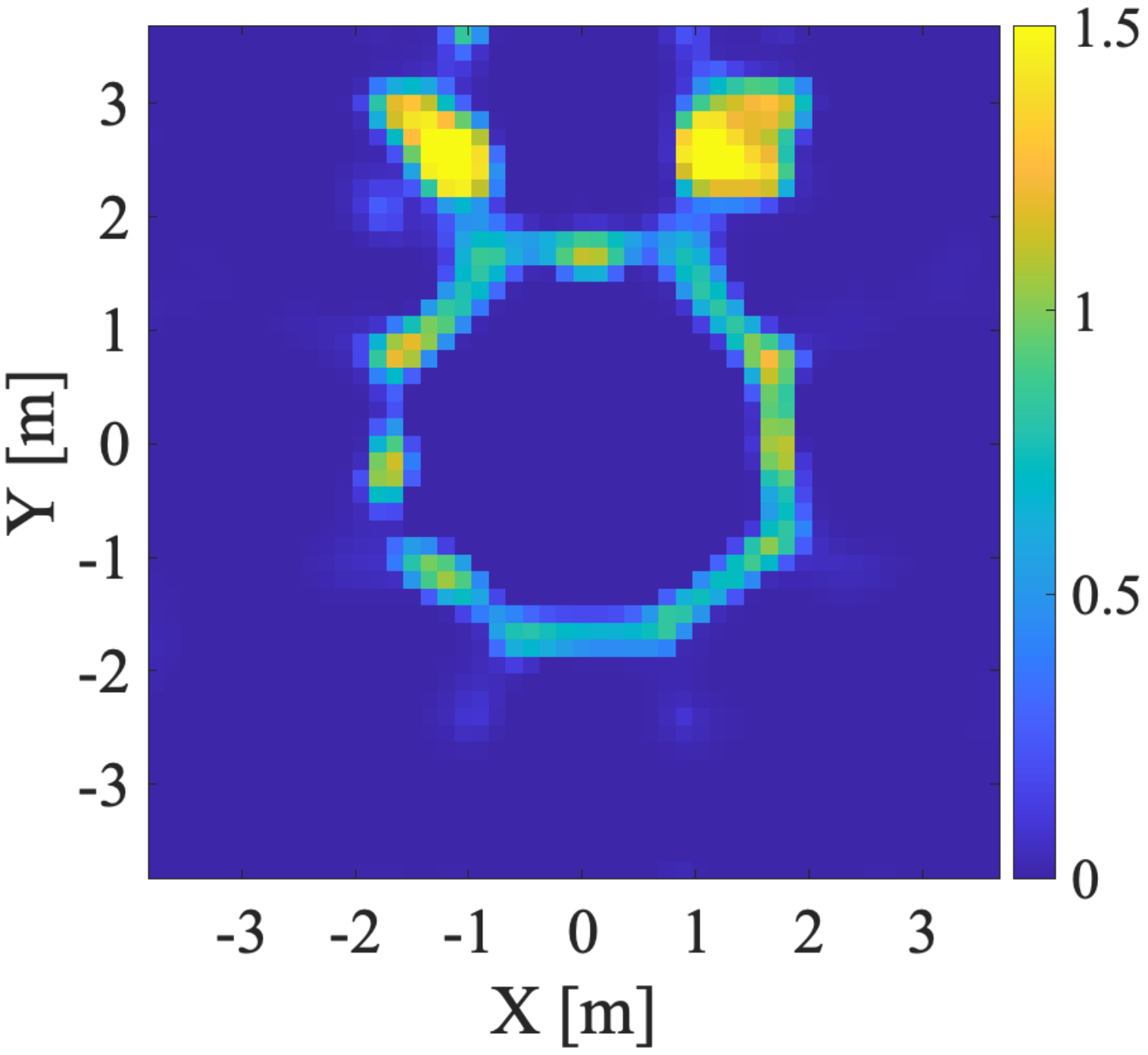}}
	\subfloat[\label{}]{\includegraphics[width=0.24\columnwidth]{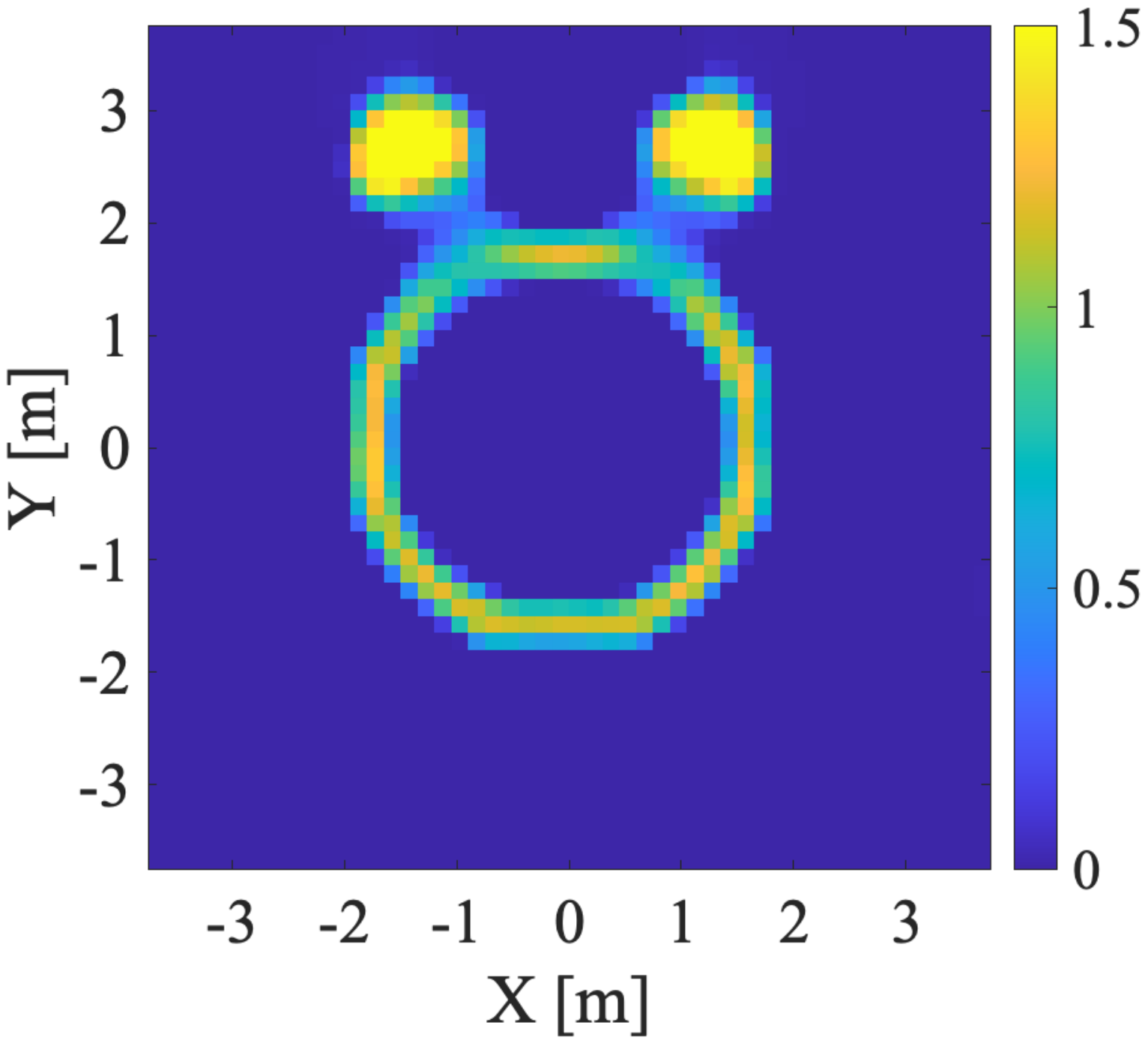}} \\
	\subfloat[\label{}]{\includegraphics[width=0.5\columnwidth]{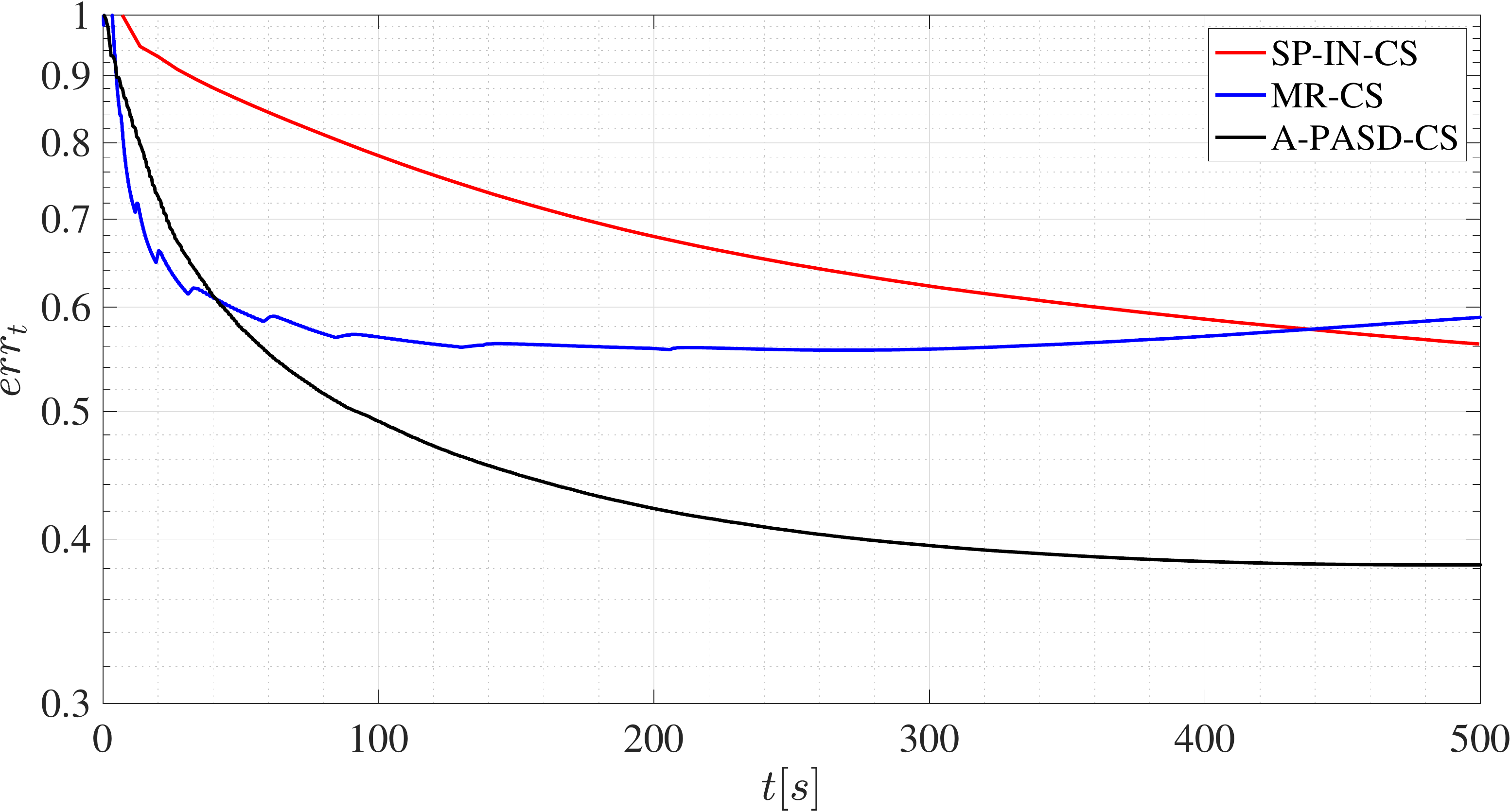}}
	\caption{(a) Investigation domain with the Austria scatterer (as represented by $\bar{\tau}^{\mathrm{ref}}$) and the transmitter and receiver 
	locations. Solutions obtained by (b) SP-IN-CS, (c) MR-CS, and (d) A-PASD-CS. (e) Reconstruction error $err_t$ versus 
	execution time $t$ for all three schemes.}
	\label{fig3}
\end{figure}

The second example is the well-known Austria profile. The ring and the two small cylinders are centered at $(0.0,0.0)$, $(2.55\,\mathrm{m},1.35\,\mathrm{m})$, and $(2.55\,\mathrm{m},-1.35\,\mathrm{m})$, respectively. The outer and inner radii of the ring are $1.95\,\mathrm{m}$ and $1.5\,\mathrm{m}$, respectively. The radii of the both small cylinders are $0.6\,\mathrm{m}$. The relative permittivities of the ring and small cylinders are $2.0$  and $2.5$, respectively. The investigation domain (as represented by ${{\bar{\tau }}^{\mathrm{ref}}}$) and transmitter-receiver configuration are shown in Fig.~\ref{fig3}(a). The sparseness level is $12.5\%$. 

Fig.~\ref{fig3} (b)-(d) show ${\bar \tau}_{t=506 \, {\mathrm{s}}}$, ${\bar \tau}_{t=501 \, {\mathrm{s}}}$, ${\bar \tau}_{t=501 \, {\mathrm{s}}}$ obtained using SP-IN-CS, MR-CS, and A-PASD-CS, respectively. The corresponding error levels are $err_{t=506 \, {\mathrm{s}}}=56\%$, $err_{t=501 \, {\mathrm{s}}}=59\%$ and $err_{t=501 \, {\mathrm{s}}}=39\%$. Note that the number of iterations required by SP-IN-CS, MR-CS, and A-PASD-CS to produce these images is $75$, $9100$, and $5293$, respectively. Fig. \ref{fig3}(e) plots $err_t$ versus execution time $t$ for all three schemes. These results show that A-PASD-CS converges faster than the other two schemes.  

\begin{figure*}[!t]
	\centering
	\subfloat[\label{}]{\includegraphics[width=0.24\columnwidth]{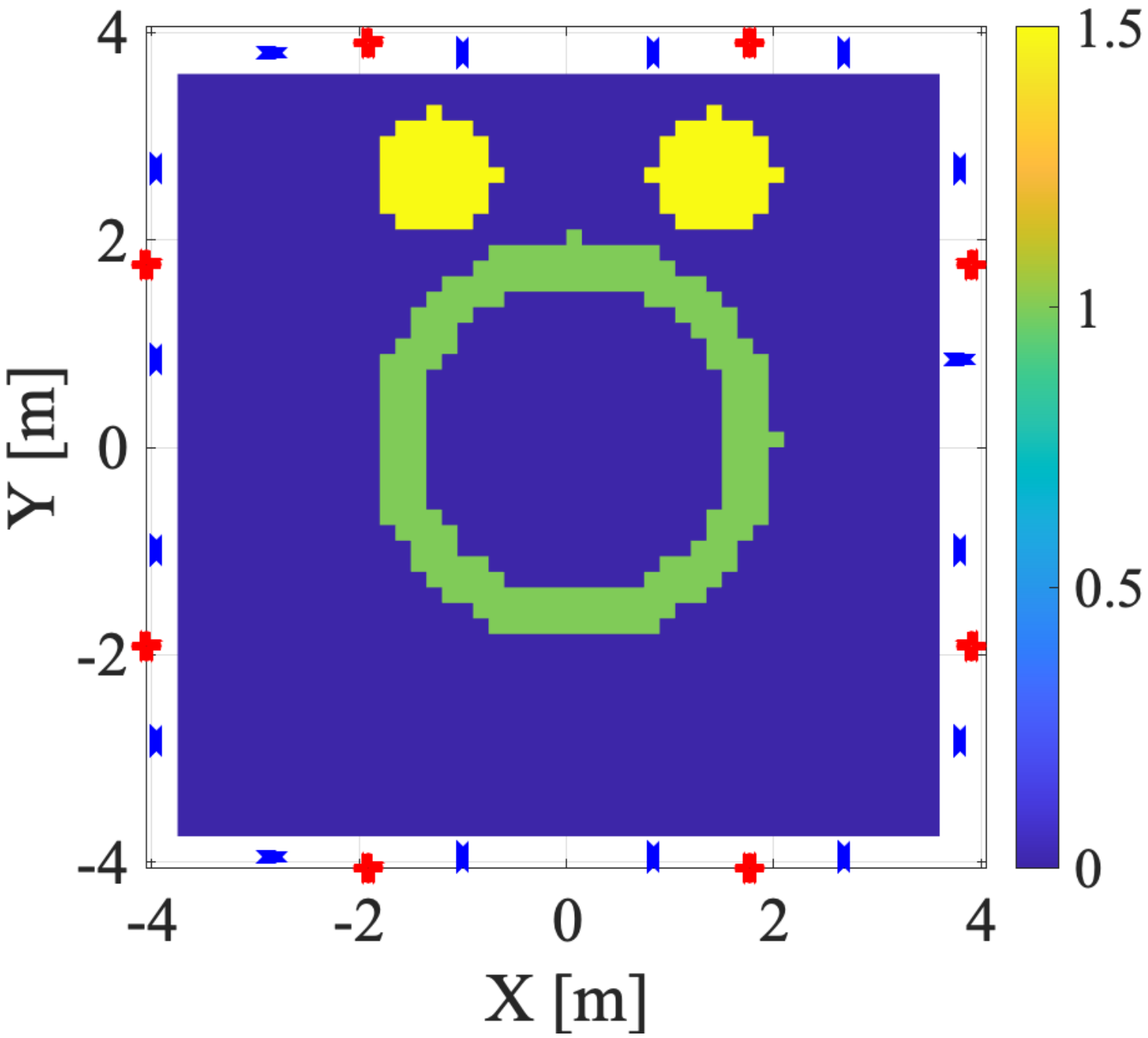}}
	\subfloat[\label{}]{\includegraphics[width=0.25\columnwidth]{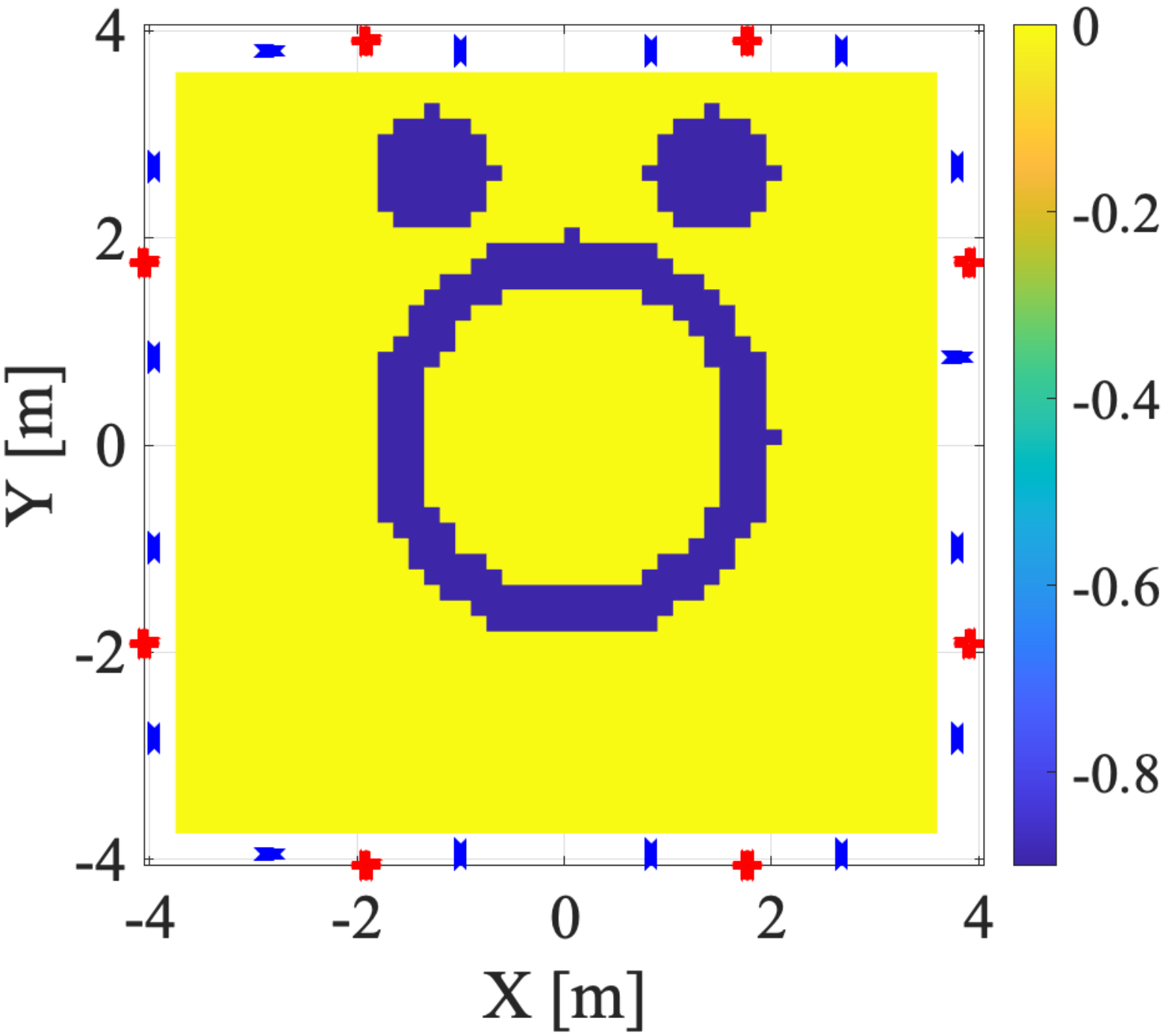}}
	\subfloat[\label{}]{\includegraphics[width=0.24\columnwidth]{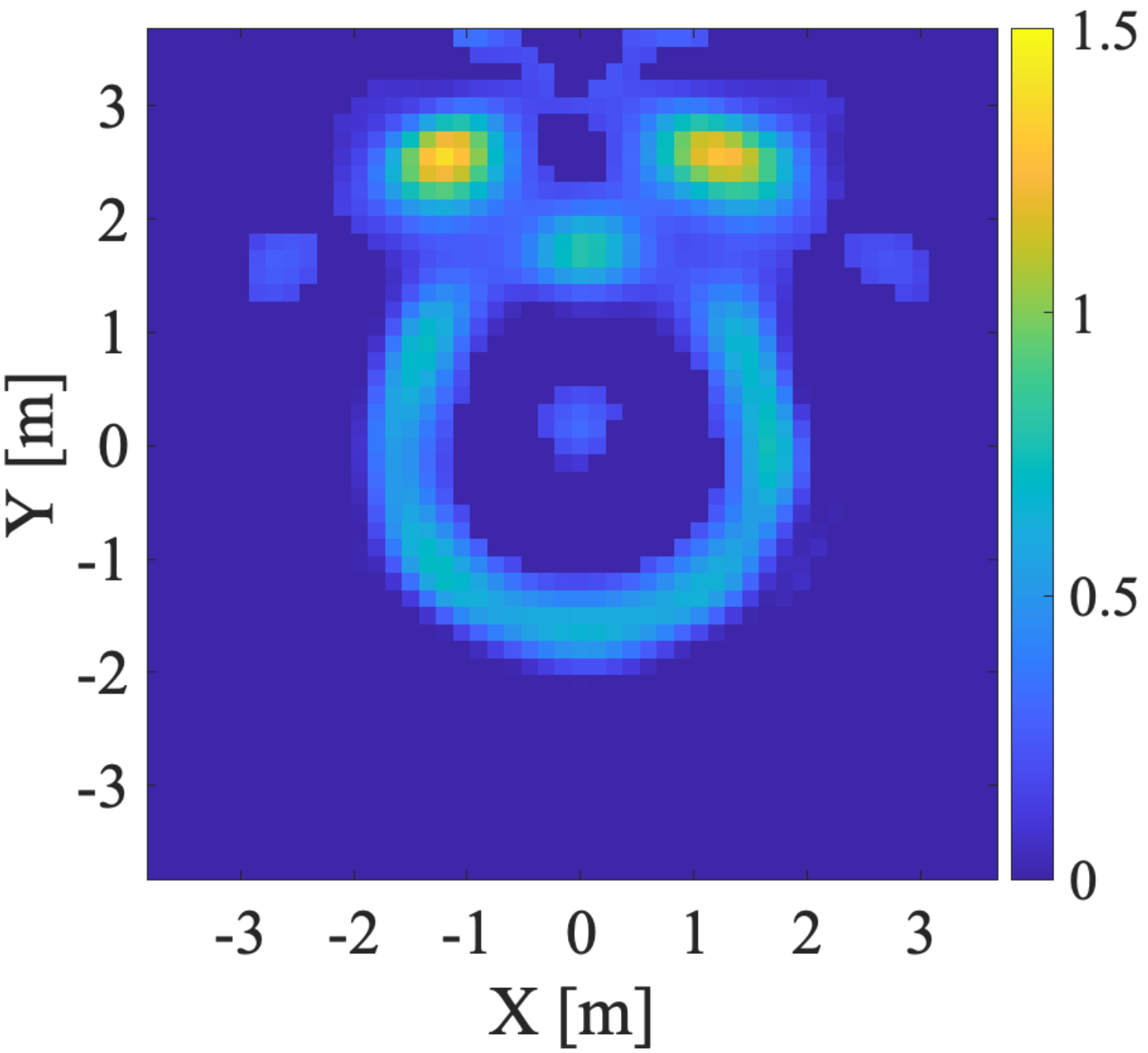}}
	\subfloat[\label{}]{\includegraphics[width=0.24\columnwidth]{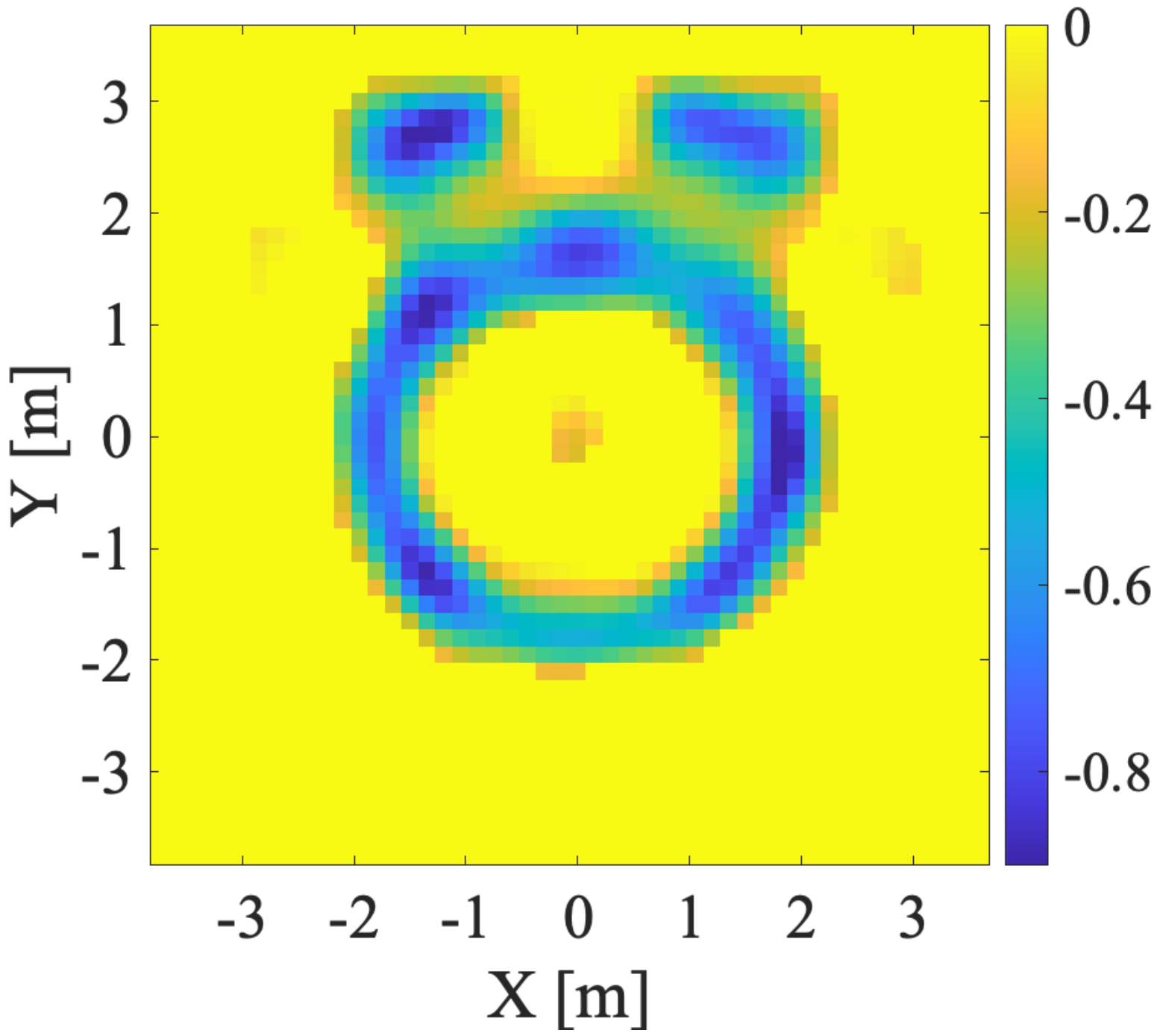}} \\
	\subfloat[\label{}]{\includegraphics[width=0.24\columnwidth]{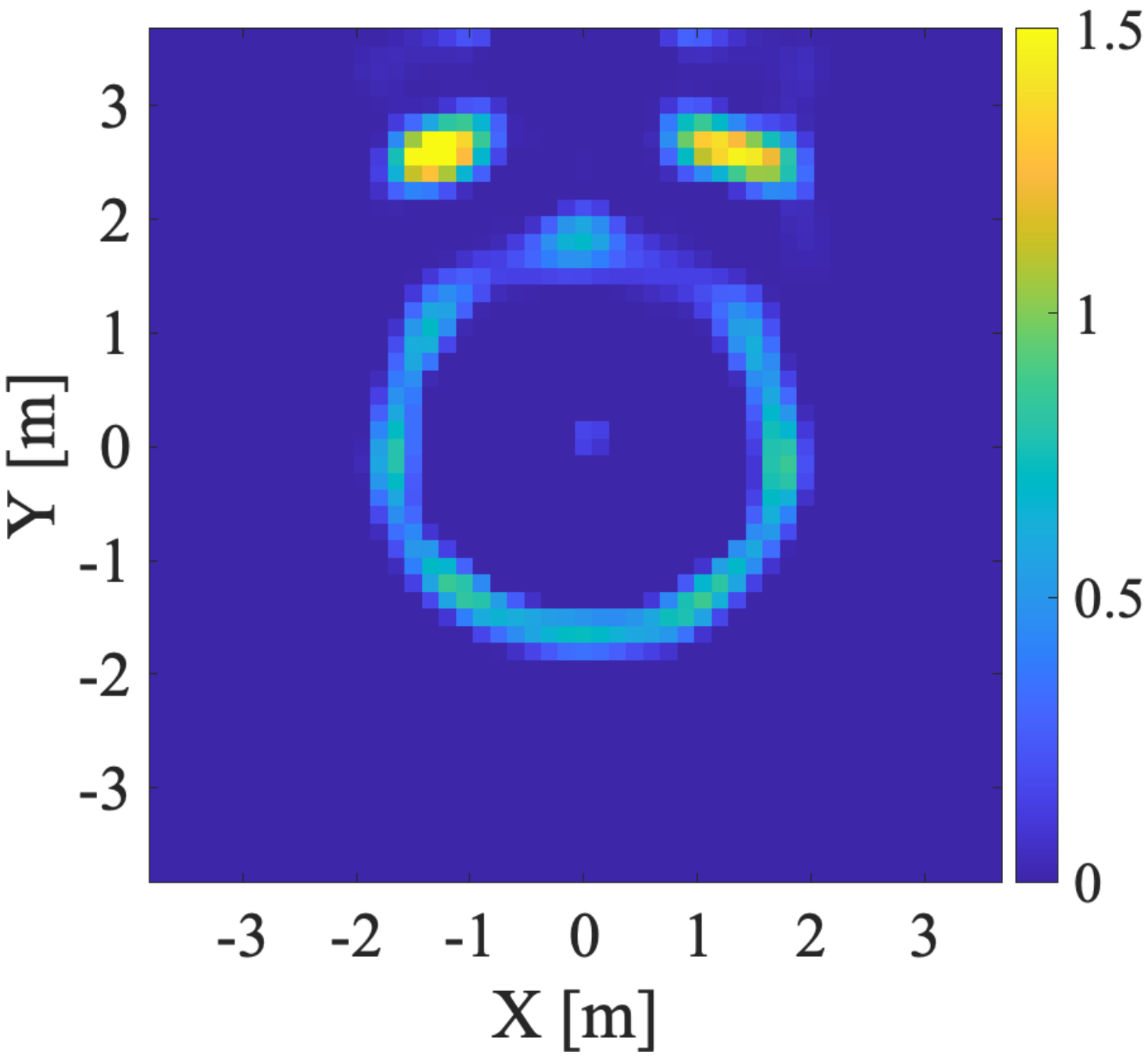}}
	\subfloat[\label{}]{\includegraphics[width=0.24\columnwidth]{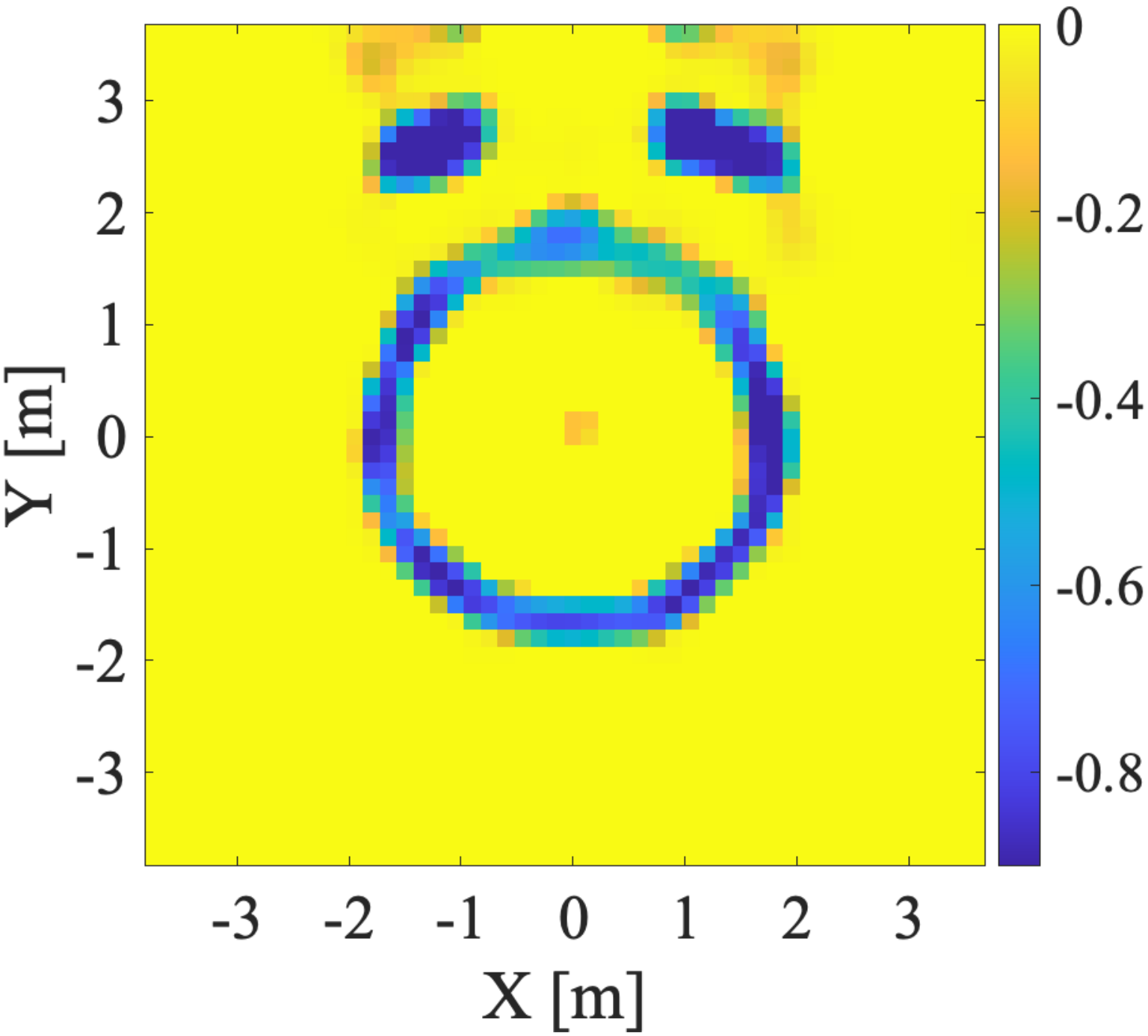}}
	\subfloat[\label{}]{\includegraphics[width=0.24\columnwidth]{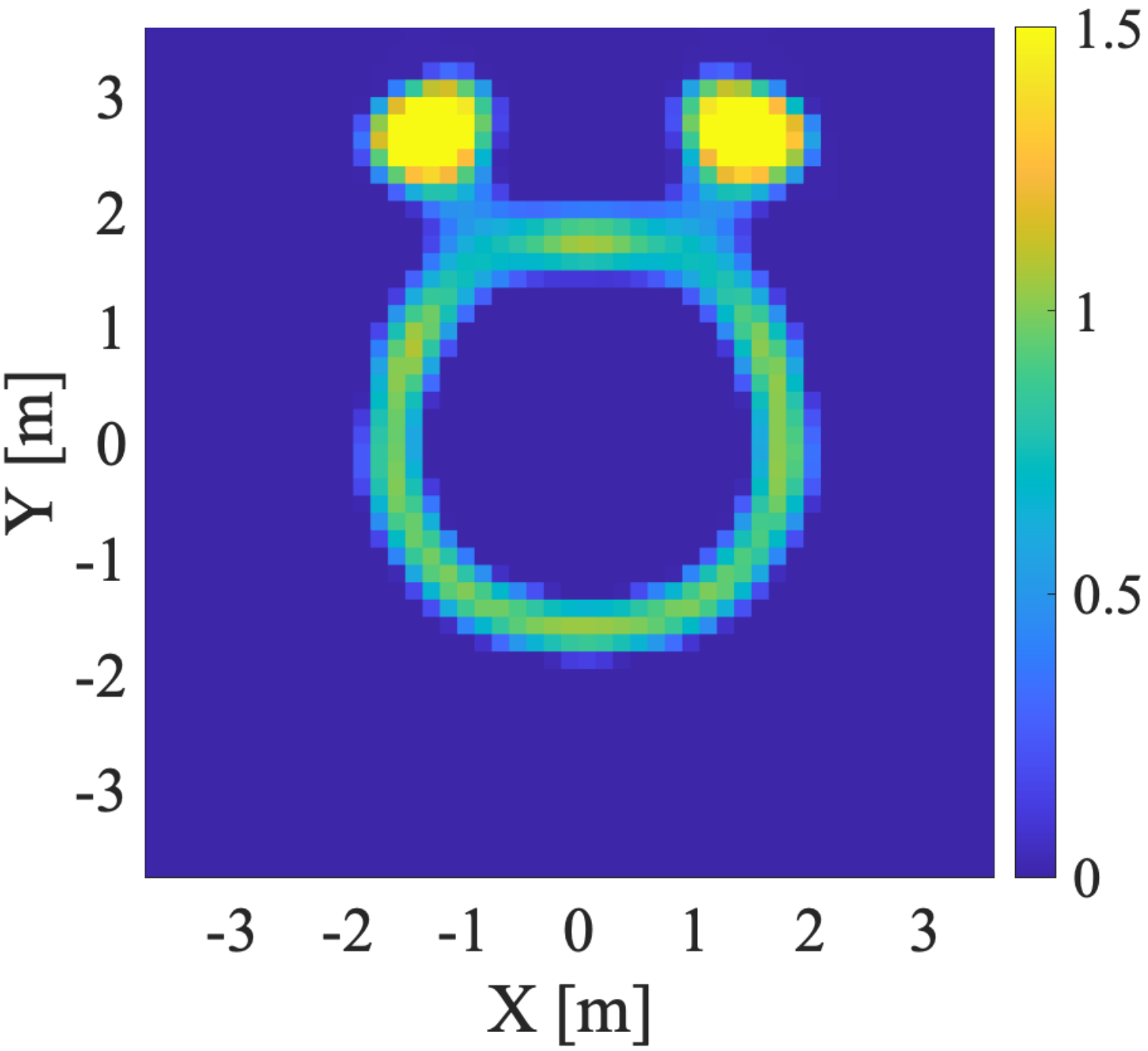}}
	\subfloat[\label{}]{\includegraphics[width=0.25\columnwidth]{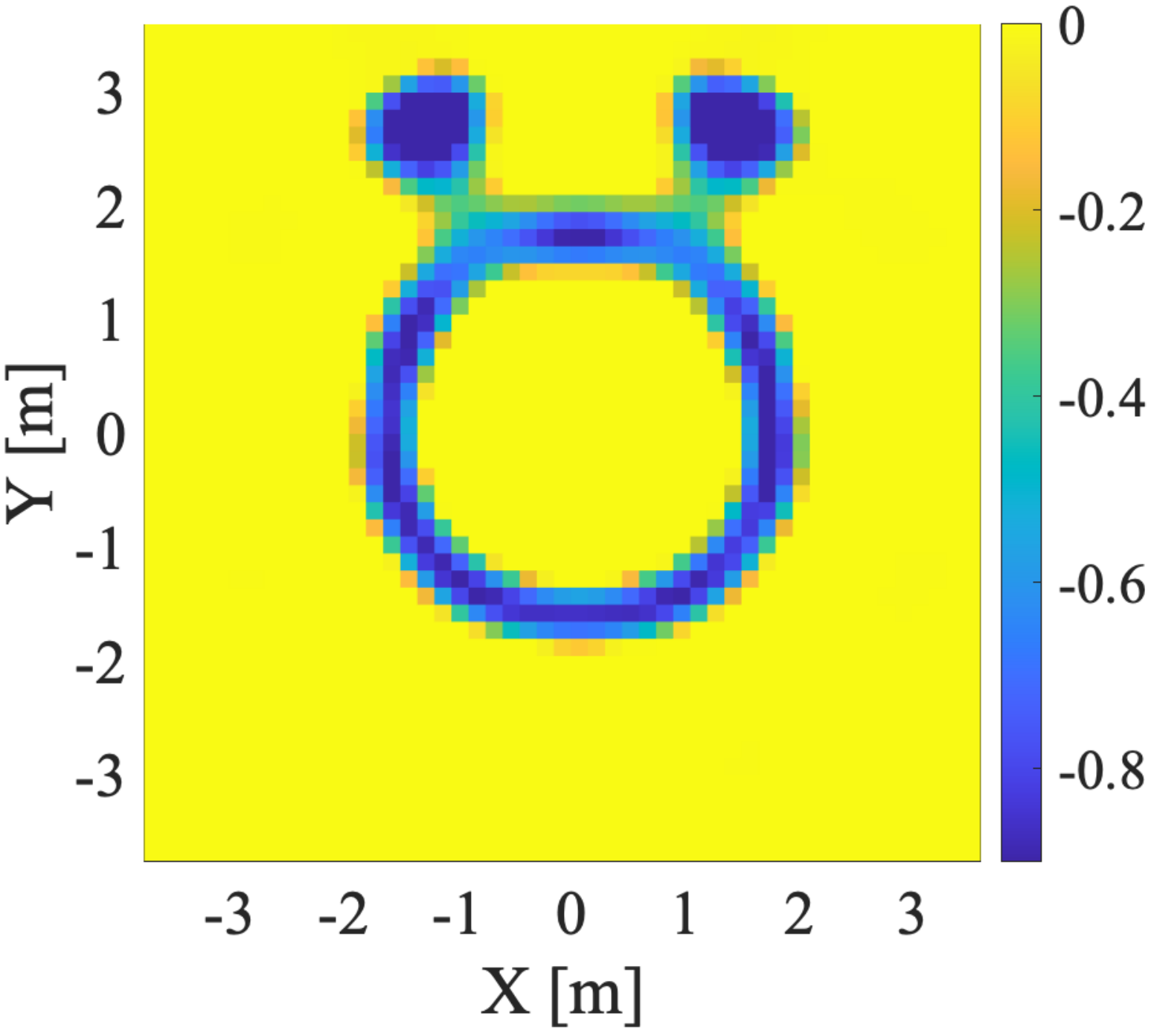}} \\
	\subfloat[\label{}]{\includegraphics[width=0.49\columnwidth]{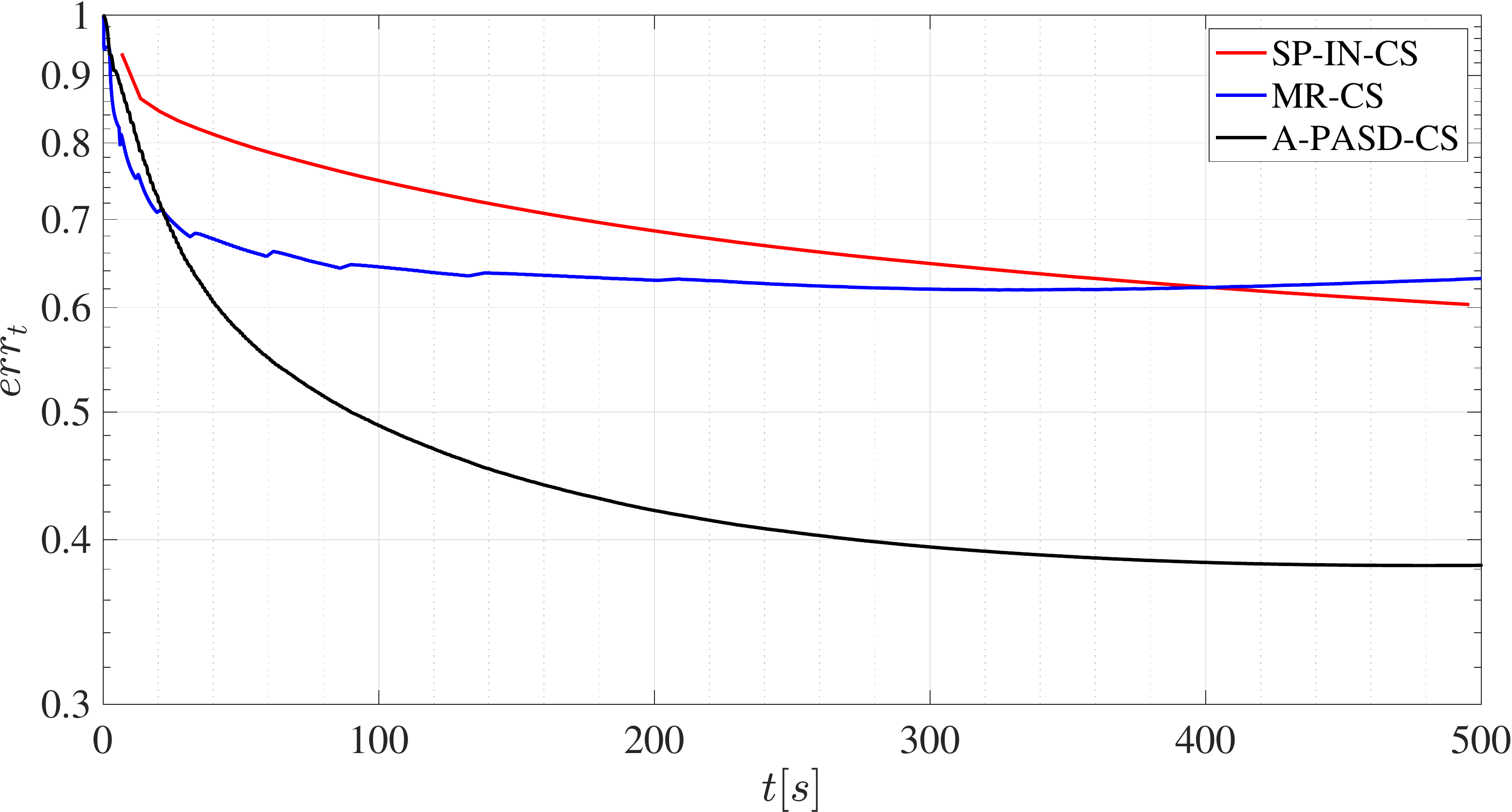}}
	\subfloat[\label{}]{\includegraphics[width=0.49\columnwidth]{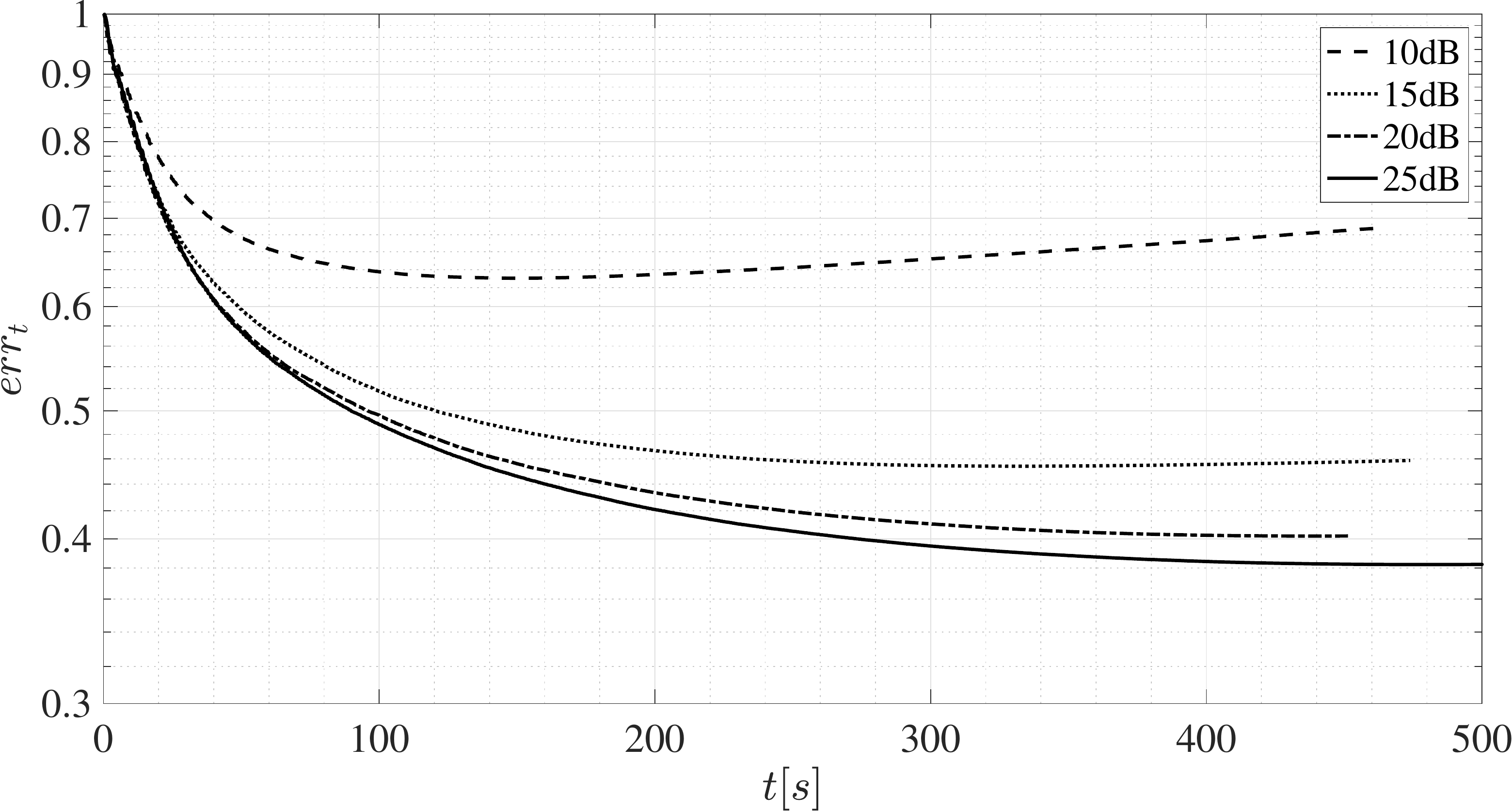}}
	\caption{ Investigation domain with the lossy Austria scatterer as represented by (a) ${\mathrm {Re}}\{{\bar \tau}^{\mathrm {ref}}\}$ and (b) ${\mathrm{Im}}\{{\bar\tau}^{\mathrm{ref}}\}$. (c) Real part and (d) imaginary part of the solution obtained by SP-IN-CS. (e) Real part and (f) imaginary part of the solution obtained by MR-CS. (g) Real part and (h) imaginary part of the solution obtained by A-PASD-CS. (i) Reconstruction error $err_t$ versus execution time $t$ for the three schemes. (j) Reconstruction error $err_t$ versus execution time $t$ for A-PASD-CS with four different levels of noise.}
\label{fig4}
\end{figure*}

Next, the contrast of the Austria scatterer is made complex by introducing a conductivity of $5\, \mathrm{m/S}$. All other configuration parameters are kept the same. Fig. \ref{fig4} (a) and (b) show ${\mathrm{Re}}\{{\bar \tau}^{\mathrm{ref}}\}$ and ${\mathrm{Im}}\{{\bar \tau}^{\mathrm{ref}}\}$, respectively. Fig.\ref{fig4} (c)-(d), (e)-(f), and (g)-(h) show ${\mathrm{Re}}\{{\bar \tau}_{t=503 \, {\mathrm{s}}}\}$-${\mathrm{Im}}\{{\bar \tau}_{t=503 \, {\mathrm{s}}}\}$, ${\mathrm{Re}}\{{\bar \tau}_{t=501 \, {\mathrm{s}}}\}$-${\mathrm{Im}}\{{\bar \tau}_{t=501 \, {\mathrm{s}}}\}$, and ${\mathrm{Re}}\{{\bar \tau}_{t=501 \, {\mathrm{s}}}\}$-${\mathrm{Im}}\{{\bar \tau}_{t=501 \, {\mathrm{s}}}\}$ obtained using SP-IN-CS, MR-CS, and A-PASD-CS, respectively. Corresponding error levels are $err_{t=503 \, {\mathrm{s}}}=60\%$, $err_{t=501 \, {\mathrm{s}}}=63\%$ and $err_{t=501 \, {\mathrm{s}}}=40\%$. Note that the number of iterations required by SP-IN-CS, MR-CS, and A-PASD-CS to produce these images is $74$, $8800$, and $5302$, respectively. Fig.~\ref{fig4}(f) plots $err_t$ versus execution time $t$ for all three schemes. These results show that A-PASD-CS indeed maintains its convergence rate and accuracy for investigation domains with complex contrast.
Next, the effect of level of the noise in the measured fields $E_{i}^{\mathrm{mea}}(\mathbf{r}_{m}^{\mathrm{R}})$ on the accuracy of the A-PASD-CS scheme is demonstrated. The reconstruction of the lossy Austria profile is carried out for three additional noise levels of $20\,\mathrm{dB}$, $15\,\mathrm{dB}$, and $10\,\mathrm{dB}$ (in addition to $25\,\mathrm{dB}$). Fig.\ref{fig4}(j) compares $err_t$ versus execution time $t$ for all four values of noise level. The accuracy and convergence of the solution deteriorate significantly when the noise level reaches $10\,\mathrm{dB}$.

\section{Conclusion}
The A-PASD algorithm and the CS formulation are incorporated and the resulting scheme is used for solving the nonlinear EM  inverse scattering problem on sparse 2D investigation domains. This scheme benefits from the faster convergence rate of the A-PASD and lower computational cost of the CS formulation. Numerical results demonstrate that the proposed algorithm is faster than existing CS-based inversion schemes. The current implementation of the A-PASD-CS scheme requires tuning of several parameters. Development of a scheme, which relies on artificial neural networks trained with various scattering scenarios to estimate these parameters, is underway.


\newpage

\begin{table}
\centering
\caption{Parameters of the A-PASD-CS scheme used for the numerical examples.} 
\label{tab1}
\vspace{0.5cm}
\begin{tabular}{|c|c|c|c|c|c|c|c|cl} \hline 
& $\alpha$ & $\psi$ & $\delta$ & $\mu$ & $\rho$ & $\lambda_0$ \\ \hline 
$\mathrm{Coaxial} $ & 0.0824 & $0.02$ & 0.2 & 0.75& 0.8 & 0.25  \\ \hline 
$\mathrm{Austria} $ & 0.0491& $0.02$ & 0.25 & 0.5 & 0.8 & 0.25  \\ \hline 
$\mathrm{Lossy Austria} $ & 0.0491 & $0.02$ & 0.25 & 0.5 & 0.8 & 0.25 \\ \hline
\end{tabular}
\end{table}

\end{document}